\documentclass{article}[11]
\usepackage{fullpage}
\usepackage{amsmath}
\usepackage{amssymb}
\usepackage{amsthm}
\usepackage{ifpdf}
\usepackage{algorithmic}
\usepackage{subfig}
\usepackage{wrapfig}
\usepackage{cite}

\usepackage{commands-tam}

\ifpdf

  \usepackage[pdftex]{epsfig}
  \usepackage[pdftex]{hyperref}

\else

    \usepackage[dvips]{epsfig}
    \newcommand{\href}[2]{#2}

\fi

\theoremstyle{definition}
\newtheorem{theorem}{Theorem}[section]

\title{\textbf{Asynchronous Signal Passing for Tile Self-Assembly:} Fuel Efficient Computation and Efficient Assembly of Shapes}
\author{
  Jennifer E. Padilla
   \thanks{Department of Chemistry, New York University,
     \protect\url{jp164@nyu.edu} This author's research was supported by National Science Foundation Grant CCF-1117210.}
\and
  Matthew J. Patitz%
    \thanks{Department of Computer Science and Computer Engineering, University of Arkansas,
      \protect\url{mpatitz@self-assembly.net} This author's research was supported in part by National Science Foundation Grant CCF-1117672.}
\and
  Raul Pena
   \thanks{Dept of Computer Science, University of Texas--Pan American,
     \protect\url{nb-raul@utpa.edu}. This author's research was supported in part by National Science Foundation Grant CCF-1117672.}
\and
  Robert T. Schweller%
    \thanks{Department of Computer Science, University of Texas--Pan American,
      \protect\url{schwellerr@cs.panam.edu} This author's research was supported in part by National Science Foundation Grant CCF-1117672.}
\and
  Nadrian C. Seeman
   \thanks{Department of Chemistry, New York University,
     \protect\url{ned.seeman@nyu.edu} This author's research was supported by National Science Foundation Grant CCF-1117210.}
\and
  Robert Sheline
   \thanks{Dept of Computer Science, University of Texas--Pan American,
      \protect\url{b.sheline@utpa.edu}. This author's research was supported in part by National Science Foundation Grant CCF-1117672.}
\and
  Scott M. Summers
   \thanks{Department of Computer Science, University of Wisconsin--Oshkosh,
     \protect\url{summerss@uwosh.edu}}
\and
  Xingsi Zhong
   \thanks{Department of Computer Science, Clemson University,
      \protect\url{zhongxingsi@gmail.com}. This author's research was supported in part by National Science Foundation Grant CCF-1117672.}
}
\date{}

\begin{document}

\maketitle

\begin{abstract}
In this paper we demonstrate the power of a model of tile self-assembly based on active glues which can dynamically change state.
We formulate
the Signal-passing Tile Assembly Model (STAM), based on the model of Padilla, et al.\cite{PadillaLiuSeeman} to be asynchronous, allowing any action of turning a
glue on or off, attaching a new tile, or breaking apart an assembly to happen in any order.
Within this highly generalized model
we provide three new solutions to tile self-assembly problems that have been addressed within the abstract Tile Assembly Model and
its variants,
showing that signal passing tiles allow for substantial improvement across multiple complexity metrics.
Our first result utilizes a recursive assembly process to achieve \emph{tile-type efficient} assembly of linear structures,
using provably fewer tile types than what is possible in standard tile assembly models.
Our second system of signal-passing tiles simulates any Turing machine with high \emph{fuel efficiency} by using only a constant number of tiles per computation step.
Our third system assembles the discrete Sierpinski triangle, demonstrating that this pattern can be \emph{strictly self-assembled} within the STAM.
This result is of particular interest in that it is known that this pattern cannot self-assemble within a number of well studied tile self-assembly models.
Notably, all of our constructions are at temperature 1, further demonstrating that signal-passing confers the power
to bypass many restrictions found in standard tile assembly models.
\end{abstract}

\section{Introduction}

The abstract Tile Assembly Model (aTAM) \cite{Winf98} created a paradigm for computation to be carried out by a physical assembly process that
captured the essence of the Wang tiling model \cite{wang_proving_1961}. Turing machine simulation within the aTAM demonstrated its
capacity for universal computation, and many subsequent assembly programs shifted focus to patterns, shapes and
structures as the output of tile computation \cite{RotWin00, RoPaWi04, soloveichik_complexity_2007, kao_randomized_2008}.
Many modifications to the standard aTAM have been investigated, including several variants that capture the notion of hierarchical assembly
 \cite{AGKS05g, DDFIRSS07, becker_pictures_2009, Versus, CheDot12, PadillaLiuSeeman}. %
Previous work introduced the notion of using signaled glue activation \cite{PadillaLiuSeeman, majumder_activatable_2007},
in particular to guide hierarchical assembly, enabling recursive self assembly
\cite{PadillaLiuSeeman}. %
Here we develop a general model of signaled tile self-assembly to
enrich the tile assembly paradigm with greater capabilities that anticipate advancing techniques in the field of DNA nanotechnology and allow tile
assembly to more closely emulate biological and natural processes.

Signaled glue activation was introduced in \cite{PadillaLiuSeeman} for the purpose of establishing communication inside an assembly so that by
activating glues at the exterior of the assembly it may take on new identities or roles.
Interactions between assemblies, as described in hierarchical models such as the 2-HAM
\cite{adleman_combinatorial_2002, AGKS05g},
can simulate the interactions of individual tiles, coordinated in the STAM by the introduction of signals.
In particular, recursive assembly results when supertiles simulate the original tiles of the tile set \cite{PadillaLiuSeeman}, a strategy
outlined here in section 4 %
in a scheme for efficient line assembly.

Cooperativity, where more than one tile must be in place to determine which tile may be added next,
is an important aspect of tile assembly systems.
A binding threshold is included in the aTAM, given as the temperature, $\tau$  of the system.
Tiles
can bind only if the sum of glue interactions at a particular site meet or exceed $\tau$, thus a
system at temperature 2
readily includes cooperativity, whereas it is not as readily achieved at temperature 1.
In the STAM,
cooperativity can occur at temperature 1 via a query process, where a tile binds to an assembly at one
edge, then queries a neighboring tile by turning on a set of glues. Information about the neighboring tile is gained
based on which of these glues binds to its match. This cooperative effect differs from the aTAM in that STAM tiles
may also respond to the identities and binding events of more distant tiles,
enabling the constructions given in this paper to operate at temperature 1.
Though the constructions here do not demonstrate a full simulation of the aTAM
at temperature 2 by the STAM at temperature 1, the results here are significant given the known and conjectured
limitations to temperature 1 computation in the aTAM and its variants \cite{CookFuSch11,jLSAT1}.

We expect glue deactivation to be as easy to implement as glue activation based on
plausible DNA strand displacement mechanisms. Therefore, we utilize this new
capability to design a Turing Machine that is fuel efficient (Section ~\ref{sec:TM}),
and to implement the strict self-assembly of the Sierpinski triangle (Section ~\ref{sec:triangle}).
While on first consideration, glue deactivation
might be thought to be on par with negative glues, it never requires repulsive forces between tiles, a necessity for negative glues that to our knowledge has yet to be implemented.
Glue deactivation serves here to produce a fuel efficient Turing Machine that does not need to
copy the state of unchanged positions on the tape.
A halting computation produces
an output tape, not a transcript of the computation as in the traditional aTAM simulation of a Turing Machine.
Strict self assembly of the Sierpinski triangle is achieved by releasing tiles that are not part of the target structure.

The addition of signaled glue activation and deactivation to tile assembly brings it one step closer to emulating biological processes of self assembly,
where communication
within a developing and growing living organism are crucial to its survival and success.
In this construction, it becomes easier to view the Turing Machine plus its tape as a developing entity, that by following its
input instructions, much as a cell follows its genetic recipe, goes through a metamorphosis and emerges from a halting computation as a new entity.
The asynchronous growth process of the strict Sierpinski triangle in this model
resembles the growth of something such as coral,
where the ``living'' functional part of the system inhabits the growing frontier of the structure, laying down an enduring structure before dying and being washed away.
The constructions presented in this paper demonstrate not only a more efficient use of materials
(Table ~\ref{table:table1}) in certain cases, but also
serve to make the model more relevant in a biological context. The STAM anticipates the increasing
sophistication of molecular computation systems, as described in the next section.

\vspace{-10pt}

\begin{table*}\footnotesize
\centering
\tabcolsep=0.5\tabcolsep
\begin{tabular}{l|c|c|c|c|cl}
\textbf{$n \times 1$ Lines} & \textbf{Tiles} & \textbf{Signals} & \textbf{Temperature} & \textbf{Glue Activation} & \textbf{Glue Deactivation}
\\ \hline
aTAM & n & - & 1 & - & -
\\ \hline
STAM (Thm.\ref{thm:nline}) & O(1) & $O(\log n)$ & 1 & Yes. & No.
\\ \hline
\end{tabular}
\vspace{10pt}

\begin{tabular}{l|c|c|c|c|c|c|clcl}
\textbf{Turing Machine} & \textbf{Space} & \textbf{Fuel} & \textbf{Temp} & \textbf{Tiles} & \textbf{Signals} & \textbf{Glue Act.} & \textbf{Glue Deact.}
\\ \hline
aTAM (\cite{RotWin00}) & $O(\sum S_i)$ & $O(S_i)$ & 2 & $O(Q)$ & - & - & -
\\ \hline
3D aTAM (\cite{CookFuSch11}) & $O(\sum S_i)$ & $O(S_i)$ & 1 & $O(Q)$ & - & - & -
\\ \hline
Negative Glues (\cite{DotKarMasNegativeJournal}) & $O(S_i)$ & $O(S_i)$ & 2 & $O(Q)$ & - & - & -
\\ \hline
Negative Glues (\cite{SchwellerShermanSODA13}) & $O(S_i)$ & $O(1)$ & 8 & $O(Q)$ & - & - & -
\\ \hline
STAM (Thm.\ref{thm:TM}) & $O(S_i)$ & $O(1)$ & 1 & $O(Q)$ & $O(1)$ & Yes. & Yes.
\\ \hline
\end{tabular}
\vspace{10pt}

\begin{tabular}{l|c|c|c|c|c|c|cl}
\textbf{Sierpinski Triangle} & \textbf{Strict} & \textbf{Tiles} & \textbf{Signals} & \textbf{Temp} & \textbf{Scale} & \textbf{Glue Act.} & \textbf{Glue Deact.}
\\ \hline
aTAM (\cite{RoPaWi04}) & No. & 7 & - & 2 & 1 & - & -
\\ \hline
STAM (Thm.\ref{thm:strict-Sierpinski}) & Yes. & 19  & 5 & 1 & 2 & Yes. & Yes.
\\ \hline
STAM (Thm.\ref{thm:weakSierp}) & No. & $5$ & $4$ & 1 & 1 & Yes. & No.
\\ \hline
\end{tabular}

\vspace{5pt}

\caption{\footnotesize Summary of our Results in the context of previous work in the field. In the Turing machine results, $Q$ is the number of states of the Turing machine being simulated and $S_i$ is the length of the tape at step $i$ in the computation.}
\label{table:summary}
\end{table*}
\label{table:table1}

\section{Physical Basis for the Model}\label{sec:justification}

The generalized model presented here has been designed to take into consideration a DNA implementation of all aspects of signaled assembly.
We envision a physical implementation where Watson-Crick DNA base pairing provides specific glue interactions
as it has done before for DNA implementations of the
aTAM \cite{Winf98, RoPaWi04}.
Additionally, we suggest that toehold-mediated DNA strand displacement reactions \cite{yurke_dna_fuelled_2000}
may be the basis for the new elements of our model: binding-induced signaling, and glue activation or deactivation.
The physical body of a tile might be composed entirely from a DNA origami structure \cite{rothemund_folding_2006, liu_crystalline_2011} in order to provide more room
for signal pathways than the
smaller DNA structures that have been used to implement the passive tiles of the aTAM \cite{winfree_design_1998}.
Many known and tested DNA strand exchange mechanisms \cite{yin_programming_2008, zhang_dynamic_2011,
qian_simple_journal_2011}, cascades \cite{dirks_triggered_2004}, and walkers \cite{omabegho_bipedal_2009, lund_molecular_2010, yin_programming_2008, wickham_direct_2011} suggest possibilities for implementing the signal pathway, including logic gates for responding to
multiple inputs and transducers for ensuring that the activated glue can have a different sequence
from the glues on the input edges \cite{seelig_enzyme_free_2006}.
Details of a plausible
DNA origami tile construction are given in \cite{PadillaLiuSeeman}.

\begin{figure}
 \vspace{-15pt}
\begin{center}
\includegraphics[width=5in]{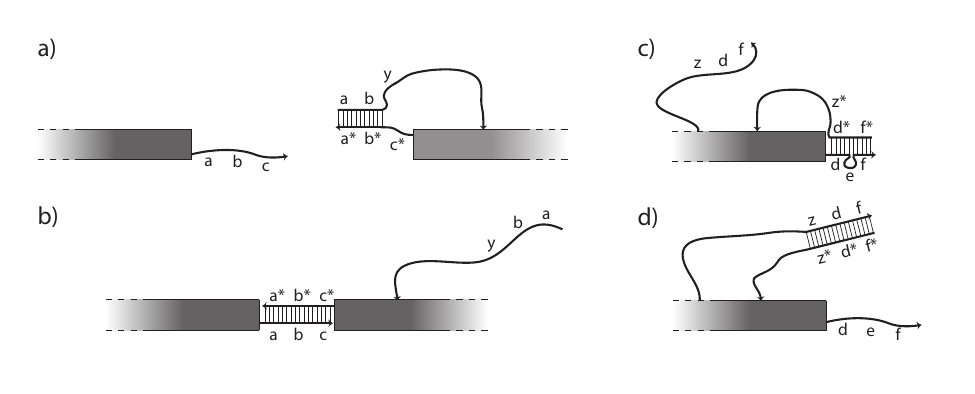}
\caption{\small DNA mechanisms for triggering a signal cascade and glue activation.
Single-stranded DNA is depicted as a line
with an arrow at the 3' end and double-stranded DNA is shown as two
parallel lines connected by a ladder representing base pairs. Letters such as $a$ and $y$
indicate short sequences of DNA bases and a star indicates the complementary sequence.
Rectangles indicate the edges of structures, possibly composed of DNA origami, to which the relevant strands of
DNA are attached. a) Complementary DNA strands of active glues can initiate binding at the
toehold $c^*$. Both glues here are in the ``$\texttt{on}$'' state because complementary toehold sequences
are exposed on both tiles.
b) Toehold binding initiates a 3-strand
DNA branch migration in which the $abc$ strand displaces the $yba$ strand as the tiles
bind to one another, freeing the $yba$ strand to trigger a signal cascade.
c) The strand $zdf$ is presumed to be unavailable until freed by a previous step in the
signal cascade (not shown). The sequence labeled $z$ may then bind at the toehold $z^*$ on the protection strand
$f^*d^*z^*$, removing it from the glue strand $def$ via strand displacement.
The glue strand $def$ is in the ``$\texttt{latent}$'' state in which the entire glue strand is protected either by
the complementary sequence on the protection strand $f^*d^*z^*$, or by inclusion in a small bulge loop,
a structure known to protect short DNA sequences from initiating base-pairing with a complementary
sequence \cite{dirks_triggered_2004}.
This scheme prevents the signal strand $zdf$ from containing a full copy of (and therefore being functionally
equivalent to) the glue strand.
d) Once the protection strand is removed, the glue strand $abc$ is active or $on$.   }
\label{fig:Binding_Activation}
\end{center}
\end{figure}

Details of our model come from a consideration of possible DNA strand displacement mechanisms that might
be used to implement signal-passing tiles.
The physical basis of the three glue states are illustrated in Figure  ~\ref{fig:Binding_Activation}, where a glue may be considered `latent' or `off' (Figure  ~\ref{fig:Binding_Activation}c) if enough of its
DNA sticky end, including the toehold, is blocked by being bound to a complementary DNA
protection strand.
Activation of a latent glue involves the toe-hold mediated removal of that strand as a result of a signal cascade
(Figure  ~\ref{fig:Binding_Activation}d).
The difference between the `on' state (Figure  ~\ref{fig:Binding_Activation}a) and the `latent' or `off'
state (Figure  ~\ref{fig:Binding_Activation}c) is based on the availability of toeholds
and whether or not there could be any opportunity for them to initiate binding to another tile in the system.
Deactivation is merely a reversal of this process, making it just as feasible as activation.

The limitation to pass a given signal only once, and to activate and deactivate a specific glue on a single
tile only once comes from the inherent consumption of ``DNA fuel'' associated with DNA strand displacement
cascades, where the production of more DNA base pairs in the products drives the process forward.
Thus, the model is designed to emphasize this fuel-consumption aspect by making each use of a signal count as
another signal path. The asynchronous nature of our model reflects the likewise asynchronous nature
of the molecular events we envision for the physical implementation. In particular,
DNA strand displacement reactions proceed via a branch migration resembling a random walk at a rate that can vary widely depending on
toehold length and composition  \cite{zhang_dynamic_2011}. We allow for the time taken by these signaling or glue activation events by storing actions triggered by various events in STAM assembly in a queue for asynchronous processing. Together, these model details capture to the best of our ability and in the most general way possible
the capabilities reported in the field of DNA nanotechnology that we see as plausible additions to the molecular
implementation of tile assembly, particularly if DNA continues to play a significant role in these implementations.

\section{STAM Notation and Model}\label{sec:model}

In this section we define the Signal-passing Tile Assembly Model (STAM), an extension of the 2-Handed Assembly Model (2HAM) \cite{AGKS05g,DDFIRSS07,Versus,CheDot12}, which itself is an extension of the Tile Assembly Model (TAM) \cite{Winf98}.  The STAM is a refined version of the model presented in \cite{PadillaLiuSeeman}, in which the basic tiles of the TAM are augmented with the ability to receive information, in the form of \emph{signals}, from neighboring tiles in an assembly, and to pass signals to neighboring tiles.  A very important feature of signals is that each signal can only move across any given tile one time - they are not reusable.

The STAM that we define is a highly generalized model which imposes minimal restrictions on aspects such as the speed of signals and orderings of events.  This generalized version, while perhaps difficult to create well-behaved constructions within, provides a framework that is intended to be maximally independent of the specific details of potential physical implementations of actual signal tiles, such as those using mechanisms suggested in \cite{PadillaLiuSeeman}.  Valid constructions within this model, such as all of the constructions presented within the following sections of this paper, will therefore also work correctly within more restricted versions of the model (for instance, where signal timing or ordering can be guaranteed).

\subsection{Informal Description of the 2HAM}
Since the STAM is an extension of the 2HAM, we now give a brief, informal description of the 2HAM.

A \emph{tile type} is a unit square with four sides, each having a \emph{glue} consisting of a \emph{label} (a finite string) and \emph{strength} (a positive integer value).  We assume a finite set $T$ of tile types, but an infinite number of copies of each tile type, each copy referred to as a \emph{tile}. A \emph{supertile} (a.k.a., \emph{assembly}) is a positioning of tiles on the integer lattice $\Z^2$. Two adjacent tiles in a supertile \emph{interact} if the glues on their abutting sides are equal (in both label and strength) and \emph{bind} with that shared strength. Each supertile induces a \emph{binding graph}, a grid graph whose vertices are tiles, with an edge between two tiles if they interact and where the weight of that edge is the strength of their bond. The supertile is \emph{$\tau$-stable} if every cut of its binding graph has strength at least $\tau$. That is, the supertile is stable if at least energy $\tau$ (i.e. a cut across bonds whose strengths sum to at least $\tau$) is required to separate the supertile into two parts.
A \emph{tile assembly system} (TAS) is a pair $\calT = (T,\tau)$, where $T$ is a finite tile set (or more generally a finite set of supertiles) and $\tau$ is the \emph{temperature}, a parameter specifying the minimum binding energy required for a supertile to be stable.  Given a TAS $\calT=(T,\tau)$, a supertile is \emph{producible} if either it is an element of $T$, or it is the $\tau$-stable result of translating two producible assemblies without overlap or rotation.
A supertile $\alpha$ is \emph{terminal} if for every producible supertile $\beta$, $\alpha$ and $\beta$ cannot be $\tau$-stably attached.
A TAS is \emph{directed} (a.k.a., \emph{deterministic}, \emph{confluent}) if it has only one terminal, producible supertile.
Given a connected shape $X \subseteq \Z^2$, a TAS $\calT$ \emph{strictly self-assembles} $X$ (also \emph{produces $X$ uniquely}) if every producible, terminal supertile places tiles exactly on those positions in $X$ (appropriately translated if necessary).  Given a pattern $Y \subseteq \Z^2$ (which must not necessarily be connected), a TAS $\calT$ \emph{weakly self-assembles} $Y$ if every producible, terminal supertile places a subset of tiles $B \subseteq T$ exactly on those positions in $Y$ (appropriately translated if necessary).  Weak self-assembly can be thought of as using a subset of tile types to ``paint a picture'' of $Y$ on a possibly larger canvas of tiles composing a terminal assembly.

\subsection{Informal Description of the STAM}
In the STAM, tiles are allowed to have sets of glues on each edge (as opposed to only one glue per side as in the TAM and 2HAM).  Tiles have an initial state in which each glue is either ``$\texttt{on}$'' or ``$\texttt{latent}$'' (i.e. can be switched $\texttt{on}$ later).  Tiles also each implement a transition function which is executed upon the binding of any glue on any edge of that tile.  The transition function specifies a set of glues (along with the sides on which those glues are located) and an action for each: 1. turn that glue $\texttt{on}$ (only valid if it is currently $\texttt{latent}$), or 2. turn that glue $\texttt{off}$ (valid if it is currently $\texttt{on}$ or $\texttt{latent}$).  This means that glues can only be $\texttt{on}$ once (although may remain so for an arbitrary amount of time or permanently), either by starting in that state or being switched $\texttt{on}$ from $\texttt{latent}$, and if they are ever switched to $\texttt{off}$ then no further transitions are allowed for that glue.  This essentially provides a single ``use'' of a glue (and thus the implicit signal sent by its activation and binding).  Note that turning a glue $\texttt{off}$ breaks any bond that the glue may have formed with a neighboring tile. Also, since tile edges can have multiple active glues, when tile edges with multiple glues are adjacent, it is assumed that all glues in the $\texttt{on}$ state bind (for a total binding strength equal to the sum of the strengths of the individually bound glues).  The transition function defined for each tile type is allowed a unique set of output actions for the binding event of each glue along its edges, meaning that the binding of any particular glue on a tile's edge can initiate a set of actions to turn an arbitrary set of the glues on the sides of the same tile either $\texttt{on}$ or $\texttt{off}$.  As the STAM is an extension of the 2HAM, binding and breaking can occur between tiles contained in pairs of arbitrarily sized supertiles.  In order to allow for physical mechanisms which implement the transition functions of tiles but are arbitrarily slower or faster than the average rates of (super)tile attachments and detachments, rather than immediately enacting the outputs of transition functions, each output action is put into a set of ``pending actions'' which includes all actions which have not yet been enacted for that glue (since it is technically possible for more than one action to have been initiated, but not yet enacted, for a particular glue). %

An STAM system consists of a set of tiles and a temperature value.  To define what is producible from such a system, we use a recursive definition of producible assemblies which starts with the initial tiles and includes any supertiles which can be formed by doing the following to any producible assembly:  1. executing any entry from the pending actions of any one glue within a tile within that supertile (and then that action is removed from the pending set), 2. binding with another supertile if they are able to form a $\tau$-stable supertile, or 3. breaking apart into two separate supertiles along a cut whose total strength is less than $\tau$.

As an overview, tiles in the STAM pass \emph{signals} to neighboring tiles simply by activating glues which can bind with glues on adjacent edges of neighboring tiles.  The information content of a signal is dependent upon the transition function of the receiving tile, that is, by what glue actions the receiving tile initiates upon the binding of its glue.  By subsequently activating and deactivating its own glues, the receiving tile can propagate the signal to any of its neighbors.  Solely by utilizing the mechanism of glue activation and deactivation initiated and carried out on individual tiles but chained together through series of glue bindings, a global network which is capable of passing information across entire assemblies (and also of allowing them to selectively enable sites for further growth or to discard arbitrary portions of the assembly), is created.  However, an important restriction is the ``fire once'' nature of the signals, meaning that each glue can only transition to any given state once, and therefore the number of signals which a tile can process is a constant dependent upon the definition of the tile type.

The STAM, as formulated, is intended to provide a model based on experimentally plausible mechanisms for glue activation and deactivation, but to abstract them in a manner which is implementation independent.  Therefore, no assumptions are made about the speed or ordering of the completion of signaling events (i.e. the execution of the transition functions which activate and deactivate glues and thus communicate with other tiles via binding events).  This provides a highly asynchronous framework in which care must be made to guarantee desired results, but which then provides robust behavior independent of the actual parameters realized by a physical system.  Furthermore, while the model allows for the placement of an arbitrary number of glues on each tile side and for each of them to signal an arbitrary number of glues on the same tile, this would obviously be limited in physical implementations.  Therefore, each system can be defined to take into account a desired threshold for each of those parameters, not exceeding it for any given tile type, and we have also defined the notion of \emph{signal complexity}, as the maximum number of glues on any side of any tile in a given set,
to capture the complexity of a tile set.

\subsection{Formal Definition of the Signal Passing Tile Assembly Model}\label{sec:STAMmodelFormal}
This section presents a formal definition of the STAM.  In order to help clarify many aspects, a consistent example is provided and referenced throughout.

\subsubsection{Basic Notation}
Let $D$ denote the set of labels $\{north, south, east, west\}$, or $\{N,S,E,W\}$.

\paragraph{\textbf{Active Glues and Glue States}}
Let $\Gamma$ denote a set of \emph{glue types}.  A \emph{glue} is an ordered pair $(g,s)$ where $g \in \Gamma$ is the glue type and $s \in \mathbb{Z}$ is the glue \emph{strength}.  Note that throughout the remainder of this paper, unless specifically stated, all glues will have strength $1$ and as shorthand will be denoted simply by their glue types with the strength omitted.

Let $Q = \{\texttt{latent}, \texttt{on}, \texttt{off}\}$ be the set of possible \emph{states} for a glue. Intuitively, $\texttt{on}$ is the ``normal'', active state of a glue, meaning that it is either able to bind or currently bound.  A glue in the $\texttt{latent}$ state is inactive (and therefore unable to bind), and also has never been $\texttt{on}$ (or $\texttt{off}$).  A glue in the $\texttt{off}$ state is also inactive and unable to bind, but can never be (re)activated. We define an \emph{active glue} as an ordered pair $(g,q)$ where $g \in \Gamma$ is a glue type and $q \in Q$ is a state.

\begin{figure}[htp]
    \vspace{-15pt}
    \begin{center}
    \includegraphics[height=0.8in]{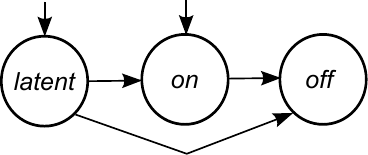} \caption{\label{fig:glue-states} \small The valid state transitions for active glues}
    \end{center}
\end{figure}

\paragraph{Active Labels}
As in the original TAM, we define a \emph{label} as an arbitrary string over some fixed alphabet and labels will be used as non-functional (i.e. they don't participate in tile bindings) means of identifying tile types. Denote as $\Sigma$ the set of all valid labels. For the self-assembly of patterns (e.g. the weak self-assembly of the Sierpinski triangle), tile labels are the mechanism used for distinguishing between groups of tiles, i.e. those ``in'' the pattern and those ``outside'' it.  Experimentally, labels can be implemented as DNA loops which protrude above the surfaces of tiles for imaging purposes \cite{liu_modifying_1999}, and thus the motivation exists to allow for the modification of tile labels along with glues.  Therefore, we define an \emph{active label} is an ordered pair $(x,q)$ where $x \in \Sigma$ and $q \in Q$.

\paragraph{Active Tiles and Transition Functions}
A \emph{generalized active tile type} $t$ is a unit square defined as a $4$-tuple $t = \left(G, L, \delta, \Pi\right)$ where $G: D \rightarrow \mathcal{P}(\Gamma \times Q)$ is a function specifying the set of all active glues present on a given side, $L$ is the set of all active labels, $\delta: D \times \Gamma \rightarrow \mathcal{P}(((D \times \Gamma \times \{\texttt{on}, \texttt{off}\}) \cup (\Sigma \times \{\texttt{on}, \texttt{off}\}))$ is the \emph{transition function} and $\Pi$ is a multiset over $(D \times \Gamma \times \{\texttt{on}, \texttt{off}\}) \cup (\Sigma \times \{\texttt{on}, \texttt{off}\})$. A generalized active tile type $t = (G, L, \delta, \Pi)$ is an \emph{active tile type} if, for all $d \in D$ and for all active glues $(g,q),(g',q') \in G(d)$, $g \ne g'$. In other words, while a tile side may have multiple glues, there cannot be multiple copies of the same type of glue on a single side.

A transition function $\delta$ takes as input a direction $d \in D$ and a glue type $g \in \Gamma$ (i.e. we say that it is ``fired'' by the binding of glue type $g$ on side $d$ of $t$), and outputs a (possibly empty) set of \emph{glue} or \emph{label actions}, i.e., elements of $(D \times \Gamma \times \{\texttt{on}, \texttt{off}\}) \cup (\Sigma \times \{\texttt{on}, \texttt{off}\})$. Consider an active tile type $t = (G, L, \delta, \Pi)$ and suppose that $(g,q) \in G(d)$ for some $d \in D$ and $(d', g' ,q') \in \delta(d,g)$. Intuitively, if $m = \Pi(d', q')$ (i.e., the \emph{multiplicity} of $(d',q') \in \Pi$) \emph{before} $\delta$ ``fires,'' then $\Pi(d', q') = m + 1$ \emph{after} executing $\delta$. We assume for the sake of convenience that $\delta$ outputs the empty set for any pair of direction and glue on which $\delta$ is not explicitly defined.  In other words, the binding of a glue on $t$ fires the transition function, which can result in a set of ``requests'' for specific active glues and labels on $t$ to transition into specified states.

As shorthand, let ``$-$'' represent ``$\texttt{latent}$'', ``$1$'' represent ``$\texttt{on}$'', and ``$0$'' represent ``$\texttt{off}$''.  Let $t = \left(G, L, \delta, \Pi\right)$ be an active tile type. For notational convenience, we will denote as $g^p_{\{q \mid \exists (d, g, q) \in \Pi\}}$ the active glue $(g, p)$ of $t$ satisfying, for some $d \in D$, $(g, p) \in G(d)$. We purposely omit the direction $d$ from our shorthand notation because it will always be clear from the context. Note that, in our notation, the superscript specifies its current state (e.g. if the active glue is $\texttt{on}$, we write $g^1$), and the subscript represents its set of \emph{pending state transitions} (i.e., the set of all glue actions $(d, g, q) \in \Pi$). For example, if $\{(N, g, \texttt{on}), (N, g, \texttt{off}), (N, g, \texttt{off})\} \subseteq \Pi$, then we write $g^p_{1,0,0}$. If there are no pending state transitions for $g$ in $\Pi$, then we omit a subscript. Figure~\ref{fig:glue-states} shows the valid state transitions for active glues, which is restricted to $\texttt{latent}$ being able to transition to either $\texttt{on}$ or $\texttt{off}$, and ``$\texttt{on}"$ being able to transition to ``$\texttt{off}"$.  Note that once a glue is in the $\texttt{off}$ state, there is no valid transition out of that state.  Also, the only valid initial states for an active glue are $\texttt{latent}$ or $\texttt{on}$ since a glue which started in the $\texttt{off}$ state could never interact and therefore could simply be removed.

\begin{figure}[htp]

    \begin{center}
    \includegraphics[height=1.6in]{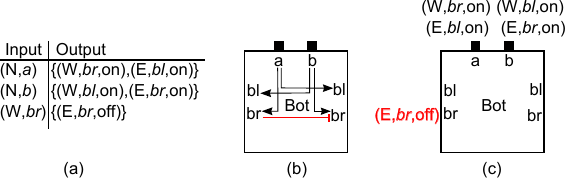} \caption{\label{fig:example-active-tile} \small An example active tile with (a) a tabular representation of its transition function, (b) a graphical depiction of its transition function using lines and arrows, and (c) a semi-graphical depiction.  In (b), an arrowed line pointing from glue $g_1$ to glue $g_2$ denotes that the binding of $g_1$ causes the glue action turning $g_2$ on to be put into $g_2$'s pending set $P$.  A red line indicates that the action is to turn the destination glue off.  In (c), the text descriptions of output actions for a given glue are placed next to its location on the tile edge.  Note that for compactness we can also combine multiple actions.  For example, if glue $a$ also had output $(W,bl,on)$ we could write $(W,(bl,br),on)$.}
    \end{center}

\end{figure}

 Figure~\ref{fig:example-active-tile} shows an example active tile type with $G(N) = \{ a^1 ,b^1 \}$, $G(W) = \{ bl^-,br^- \}$,  $G(E) = \{ bl^-,br^- \}$, $G(S) = \emptyset$ and $L = \{ \textmd{Bot}\}$.  Note that for the sake of convenience, in figures we tend to omit superscripts unless the glue state is $\texttt{off}$, so a glue with no superscript and a rectangular tab next to it represents that the glue is $\texttt{on}$, and with no such a tab is $\texttt{latent}$.  Glues which are $\texttt{off}$ will contain the ``$0$'' superscript, or be removed completely, to remove ambiguity. Figure~\ref{fig:example-active-tile} also includes two depictions of the tile's transition function.  In this example, when the glue $b$ on the North side of the tile binds, the tile's transition function specifies that glues $b_l$ on the West side and $b_r$ on the East side are requested to turn $\texttt{on}$.

An \emph{active tile} is an instance of an active tile type which has its own pending sets for each of its active glues and labels--all of which sets must be initially empty.

\paragraph{Active Supertiles}
Active supertiles in the STAM are defined similarly to supetiles in the 2HAM.  Note that only glues which are in the $\texttt{on}$ state can bind, and only those which are bound contribute to the $\tau$-stability of a supertile.  When two $\tau$-stable supertiles can be translated so that they are non-overlapping and the abutting $\texttt{on}$ glues can bind to create a cut strength of at least $\tau$, we say that they are $\tau$-\emph{combinable}. A supertile $A$ is said to be $\tau$-\emph{breakable} into supertiles $B$ and $C$ if there exists a strength $\tau' < \tau$ strength cut of the stability graph $G_A$ that separates the tiles of $A$ into supertiles $B$ and $C$.

\paragraph{Active Supertile Combination}
Two active supertiles $A$ and $B$ may combine into active supertile $C$ if the underlying supertiles for $A$ and $B$ are $\tau$-combinable into the underlying supertile for $C$.  When supertiles combine, all matched glues in the $\texttt{on}$ state on the boundary between $A$ and $B$ are said to bind, and thus fire the transition functions of their respective tiles, causing the necessary states to be added to the pending sets of the targeted tiles.

\paragraph{Active Supertile Breaking}
An active supertile $A$ can break into active supertiles $B$ and $C$ if the underlying supertile for $A$ has a cut of less than $\tau$-strength dividing $A$ into the underlying supertiles for $B$ and $C$.

An STAM system is a tuple $(T,\tau)$ where $T$ is a set of initial active supertiles referred to as the \emph{initial assembly set} of the system, and $\tau$ is a positive integer referred to as the \emph{temperature} of the system.  We further restrict that the initial assembly set only contains active super tiles $t = (G, L, \delta, \Pi)$ such that $\Pi = \emptyset$.  For some problems such as the unique assembly of shapes and lines (see section~\ref{sec:lines}), $T$ is further restricted to contain only active supertiles that consist of singleton active tiles.

\subsubsection{Producible Assemblies}
The signal passing tile assembly model is defined in terms of the set of \emph{producible} active supertiles $P_{T,\tau}$.  $P_{T,\tau}$ is defined recursively as follows:
\begin{description}
    \item[Base Set of Assemblies] $T \subseteq P_{T,\tau}$
    \item[Combination Transition] For any 2 active supertiles $A,B \in P_{T,\tau}$, if $A$ and $B$ are $\tau$-combinable into active supertile $C$, then $C\in P_{T,\tau}$.
    \item[Break Transition] For any supertile $A\in P_{T,\tau}$, if $A$ is $\tau$-breakable into active supertiles $B$ and $C$, then $B,C \in P_{T,\tau}$.
    \item[Active Glue Transition] Consider an active supertile $A\in P_{T,\tau}$. For each active tile $t \in A$ with $A(x,y) = t = (G, L, \delta, \Pi)$, for all $d \in D$, for all $(g, p) \in G(d)$ and for all $(d, g, q) \in \Pi$, there is a supertile $B \in P_{T,\tau}$ such that $A$ and $B$ differ only at location $(x,y)$ as follows: $B(x,y) = t' = (G', L, \delta, \Pi')$ satisfying
        \begin{enumerate}
            \item$\Pi' = \Pi - \{(d, g, q)\}$ and
            \item if $p = \texttt{on}$ and $q = \texttt{off}$ or $p = \texttt{latent}$ and $q = \texttt{off}$ or $p = \texttt{latent}$ and $q = \texttt{on}$, then for all $d \ne d' \in D$, $G'(d') = G(d')$, $(g, q) \in G'(d)$ and for all $q \ne q' \in Q$, $(g, q') \not \in G'(d)$.
        \end{enumerate}
        (This simply says that any supertile $B$ which can be created from another producible supertile $A$ by simply executing one of $A$'s pending actions is also producible.)
    \item[Active Label Transition] Active label transitions are defined similar to active glue transitions.
\end{description}

\subsubsection{Terminal Assemblies}
The set of producible assemblies of an STAM system defines the collection of active supertiles that can occur throughout the assembly process.  The \emph{terminal} set of assemblies $\mathcal{A}_{\Box}[T,\tau]$ of an STAM system $(T,\tau)$ defines the subset of producible assemblies for which no combination, break, glue transition, or label transition is possible.  Conceptually, the terminal assembly set represents the sink state of the assembly system in which the system has been given enough time for all assemblies to reach a terminal state.  The terminal set of assemblies is considered the output of an STAM system.  In our constructions we are interested in designing systems that will have either 1) a unique terminal assembly that has a desired shape or 2) has a collection of terminal assemblies in which the desired target assembly is the largest or 3) the desired target assembly is the only assembly which contains a specially designated \emph{marker} tile.

Please see Section~\ref{sec:STAM-example} for an example STAM system and several transitions and producible assemblies.

\subsection{Example STAM System}\label{sec:STAM-example}
Here we give a small example STAM system along with descriptions of several transitions and producible assemblies.

\subsubsection{Example tile types}
\begin{figure}[htp]
    \begin{center}
    \includegraphics[height=.8in]{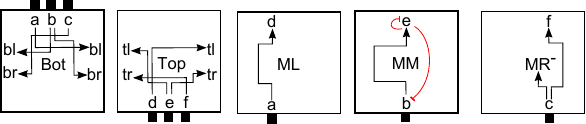} \caption{\label{fig:example-tile-set} \small An example active tile set.}
    \end{center}
\end{figure}

Figure~\ref{fig:example-tile-set} shows a tile set that will be used for the following example.  We will refer to active tiles by their labels for convenience.  The ``Bot'' tile has four glue actions defined for its transition function, all of which turn $\texttt{on}$ glues. The ``MM'' tile includes two glue actions which turn $\texttt{off}$ glues:  when glue $e$ on the North binds, both that glue and glue $b$ on the South get $\texttt{off}$ transitions placed into the $\Pi$ multiset of the active tile, which will eventually cause them to be turned $\texttt{off}$.  The transition function for the ``MR'' tile contains a label action, namely when the $c$ glue on the South binds.

\subsubsection{Example STAM transitions}
\begin{figure}[htp]
    \begin{center}
    \includegraphics[height=2.0in]{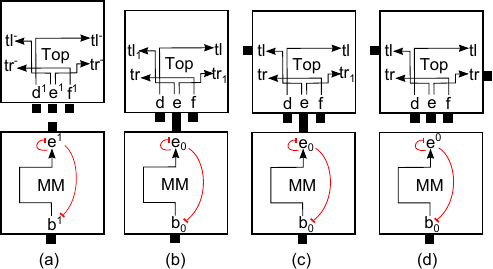} \caption{\label{fig:example-states} \small Example transitions for the tile types in Figure~\ref{fig:example-tile-set} assuming a temperature $\tau=1$ system}
    \end{center}
\end{figure}

Define an STAM system $(T,2)$ where $T$ is the tile set shown in Figure~\ref{fig:example-tile-set}.  Figure~\ref{fig:example-states} shows a set of transitions and producible assemblies within that system.  Figure~\ref{fig:example-states}(a) shows a ``Top'' and ``MM'' tile in their initial configurations (with empty $\Pi$ sets for all active tiles as required, and all superscripts explicitly shown for reference).  Figure~\ref{fig:example-states}(b) shows the supertile resulting from a combination transition performed on one ``Top'' and one ``MM'' tile.  Note that the binding of the two $e$ glues causes four glues to have state transitions added to the $\Pi$ multisets the belong to their respective active tile.  Figure~\ref{fig:example-states}(c) shows the supertile resulting from a glue flip transition performed on the supertile in Figure~\ref{fig:example-states}(b), namely the $tl$ glue on the West side of the ``Top'' tile transitioned to state $\texttt{on}$ and that pending transition was removed from the $\Pi$ multiset belonging to its active tile.  Figure~\ref{fig:example-states}(d) shows the supertile resulting after two glue flip transitions are performed on the supertile in Figure~\ref{fig:example-states}(c), namely glue $tr$ on the East of ``Top'' is turned $\texttt{on}$ and $e$ on the North of ``MM'' is turned $\texttt{off}$.

\begin{figure}[htp]
\centering
  \subfloat[][Three producible supertiles]{%
        \label{fig:example-assemblies1}%
        \centering
        \includegraphics[height=2.8in]{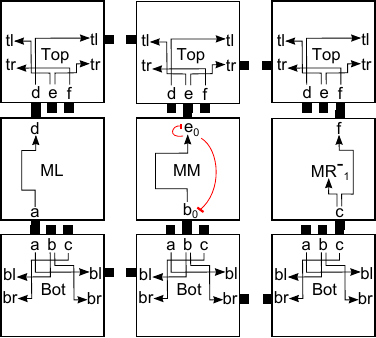}}%
        \hspace{25pt}
  \subfloat[][A single producible supertile formed by the combination of those in Figure~\ref{fig:example-assemblies1}]{%
        \label{fig:example-assemblies2}%
        \centering
        \includegraphics[height=2.8in]{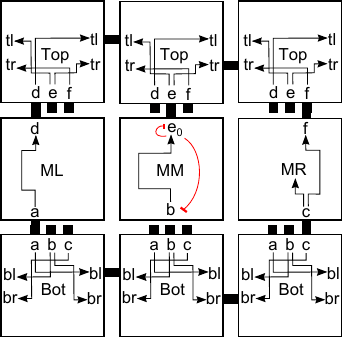}
        }%
  \centering
  \caption{Producible supertiles in the temperature $\tau=1$ system defined with the tile types in Figure~\ref{fig:example-tile-set}}
\end{figure}

\begin{figure}[htp]
    \begin{center}
    \includegraphics[height=2.0in]{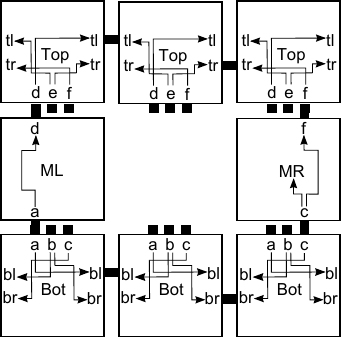} \caption{\label{fig:example-assemblies3} \small Another producible supertile in the temperature $\tau=1$ system defined with tile types from Figure~\ref{fig:example-tile-set}.  Note that it is not terminal as a new ``MM'' tile could attach in the center, repeatedly falling off and being replaced an infinite number of times.}
    \end{center}
\end{figure}

\subsection{STAM Metrics}\label{sec:metrics}
Within this paper we consider the problem of assembling precise shapes, simulating Turing machines, and the strict self-assembly of infinite fractals.  To measure how efficiently we can solve these assembly problems within the STAM model, as opposed to alternate models, we consider a number of natural metrics designed to measure experimental feasibility with respect to likely STAM implementations, the most important being DNA origami based tiles which implement glue actions through DNA strand displacement.

\subsubsection{Tile Complexity}
The tile complexity of an STAM system $(T,\tau)$ is $|T|$, the number distinct active supertiles that serve as the initial set of assemblies for the system. In the case that $T$ is restricted to contain active supertiles consisting of only singleton active tiles, this metric denotes the number of distinct tile types that must be engineered to implement the STAM system.  In general, the metric describes the number of distinct assemblies that must be engineered to implement the system.

\subsubsection{Signal Complexity}
The signal complexity of an STAM system is the maximum number of glues that occur on the face of any tile from $T$.  Within the STAM system, each glue along a tile face is potentially set up to fire a transition function that triggers a set of local glue actions upon binding.  In practice, such transition triggers can be set up through a sequence of DNA strand displacements.  Setting up a small network of such displacement chains on the face of a DNA origami tile is experimentally quite feasible.  However, the cost and complexity grow substantially as the number of distinct displacement reactions grows, making the signal complexity a key metric for experimental feasibility.

\subsubsection{Fuel Efficiency}
Fuel efficiency is a metric that is considered in the context of simulating a Turing Machine and denotes the number of tiles that are used up (and cannot be reused) per computation step of the Turing machine being simulated.  Our result constitutes the first Turing simulation self-assembly system that achieves $O(1)$ fuel efficiency.

\subsubsection{Space Complexity}
Space complexity is a metric that is considered in the context of simulating a Turing Machine and denotes the size of the assembly that represents the current tape and state of the Turing machine after a given computation step.

\subsubsection{Temperature}
For an STAM system $(T,\tau)$, the temperature value $\tau$ denotes the number of glue bonds required for two assemblies to combine.  Higher temperature systems realize finer grained bonding strength differences than lower temperature systems and may be comparably harder to implement in practice.  One of the most straightforward class of systems to implement may be temperature $\tau=1$ systems in which any single bond is sufficient to cause two assemblies to attach.  In particular, strength $\tau=1$ systems do not require the implementation of error prone \emph{cooperative bonding} in which attachment may be based on 2 or more glues spread across 2 or more distinct tiles.

\section{Efficient Construction of Linear Assemblies}\label{sec:lines}

In the aTAM, $n \times 1$ lines are inherently inefficient to self-assemble, requiring the worst possible tile complexity of $n$.  However, using signal-passing tiles it is possible to create $n \times 1$ lines using no more than $6$ tile types, regardless of the value of $n$.  Of course, the value of $n$ must some how be encoded in the system, but rather than requiring $n$ tile types as in the aTAM, in the STAM the value of $n$ can be efficiently encoded using $\log n$ bits to require $O(1)$ tile types with $O(\log n)$ signal complexity.  The construction we use
makes use of a recursive doubling strategy
where random tile binding events randomly assign the fate of each supertile at each stage, and is of independent interest in terms of using signal-passing tiles to perform recursive assembly of structures.

We first state our result for the special case where $n$ is a power of 2 and provide a detailed construction.  The exact tile set is given in Figure \ref{fig:line_tiles}.  A high-level description of the recursive doubling technique is given in Figure \ref{fig:supertileIdentity}, and a small example of the assembly process for a length $16$ line is given in Figures~\ref{fig:exampleLine1}-\ref{fig:exampleLine2}.  This result can be generalized to work for any positive integer $n$ by increasing the glues on the tile types of the powers of $2$ construction and adding $3$ more tile types, while keeping the signal complexity at $O(\log n)$, yielding the final result stated in Theorem~\ref{thm:nline}.

\begin{theorem}\label{thm:linePower}
For every $k \in \mathbb{N}$, there exists an STAM system $(T,1)$ which uniquely assembles a $2^{k} \times 1$ line. Moreover, $T$ contains $4$ tiles, has signal complexity $O(k)$, and does not use glue deactivation.
\end{theorem}

\subsection{Proof of Theorem~\ref{thm:linePower}}
\begin{proof}

Define $f(A)$ as a naming function that takes as input a supertile $A$ and returns a string $s \in  \{a_{k}, b_{k}, x_{k},y_{k} | k \in \mathbb{N}\} \cup \lambda$. For an arbitrary input $A$, if the height is 1 and the width is $2^n$ for some $n \in \mathbb{N}$, $f(A)$ returns $a_{k}$ if $A$ has active glues labeled $ax_{k}$ and $ay_{k}$ on its eastern edge,  $b_{k}$ if $A$ has active glues labeled $bx_{k}$ and $by_{k}$ on its eastern edge,  $x_{k}$ if $A$ has active glues labeled $ax_{k}$ and $bx_{k}$ on its western edge, and $y_{k}$ if $A$ has active glues labeled $ay_{k}$ and $by_{k}$ on its western edge. Otherwise, $f(A)$ returns $\lambda$.

Initially, the system consists only of copies of tiles from $T$, namely the four singleton tiles whose names correspond to their identities as defined by $f$: $a_{0}, b_{0}, x_{0},$ and $y_{0}$.  Assembly proceeds by recursive combination of these tiles into larger supertiles. Figure ~\ref{fig:supertileIdentity} shows how producible assemblies can join; for example, $a_{0}$ can attach to the western edge of either $x_{0}$ or $y_{0}$, producing $a_{1}$ and $y_{1}$, respectively. For a detailed view of the internal signal structure of the tiles that allows for this behavior, see Figure~\ref{fig:line_tiles}.  Note that each assembly possesses two exposed external glues which allow non-deterministic binding to occur, a critical aspect of the construction.

The binding event that joins two assemblies together will propagate a signal either left or right (via $L_k$ or $R_k$ glues), depending on what identity the newly formed assembly is supposed to take on. For example: if tile $a_{0}$ attaches to tile $x_{0}$, it can only do so by virtue of glue $ax_{0}$. At this point, the assembly ``knows'' that it is now supposed to be $a_{1}$, and activates a glue on its right face: $R_{1}$. Each tile in the assembly can receive $R_{1}$ on its left edge (because it is already activated), and activate it on its right edge, effectively propagating a signal across multiple tiles. When the rightmost tile (in this case, $x_{0}$) receives the signal, it activates glues corresponding to the appropriate identity, i.e., $ax_{1}$ and $ay_{1}$. At this point, the assembly can properly be called $a_{1}$. The tileset is designed in such a way that the external edge tiles in any particular assembly are known. For example, it can be observed that the leftmost and rightmost tiles in any $a_{k}$ will be $a_{0}$ and $x_{0}$, respectively. This fact allows a simple signal transduction: when an edge tile receives a particular signal, there is exactly one identity that particular assembly can adopt via glue activation.

The assembly process can be thought of as being isomorphic to a staged assembly process, where at every stage $k$ such that $0 < k \le \log_{2}n$, four new assemblies are formed from assemblies produced at stage $k -1$. These can combine with one another in four ways, and so on. This process terminates at the final mixing, when $k=\log_{2}n$, because the final binding event does not trigger signal propagation, resulting in terminal $n \times 1$ lines.  Note that there are $4$ unique terminal assemblies, all of which are $n \times 1$ lines.

\begin{figure}
\begin{center}
\includegraphics[width=0.98\linewidth]{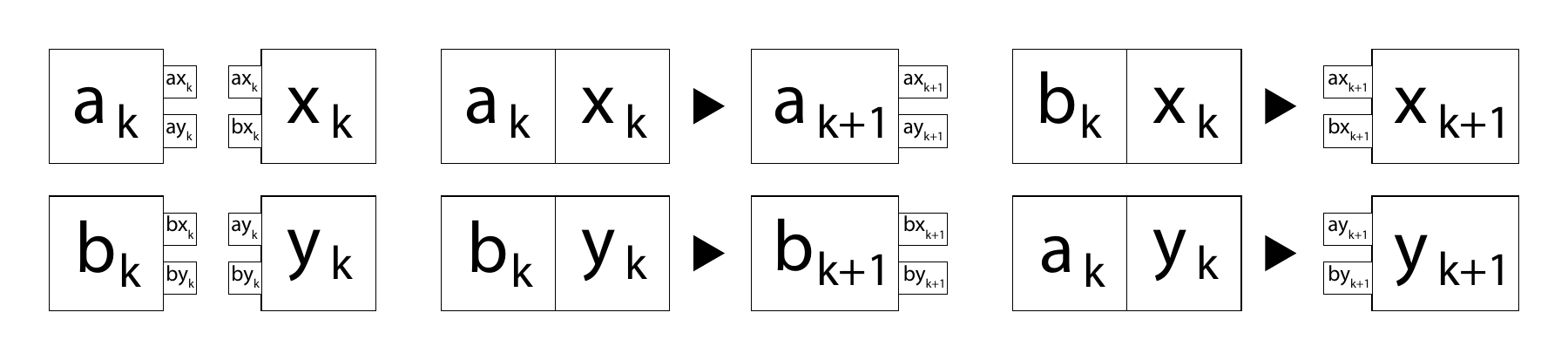}
\caption{\scriptsize Four recursively defined types of assemblies: $a_{k}, b_{k}, x_{k}, y_{k}$. For example: $a_{k}$ is composed of $a_{k-1}$ and $x_{k-1}$ and has glues $ax_{k}$ and $ay_{k}$ activated on its eastern edge.}
\label{fig:supertileIdentity}
\end{center}
\end{figure}

\begin{figure}
\begin{center}
\includegraphics[width=0.98\linewidth]{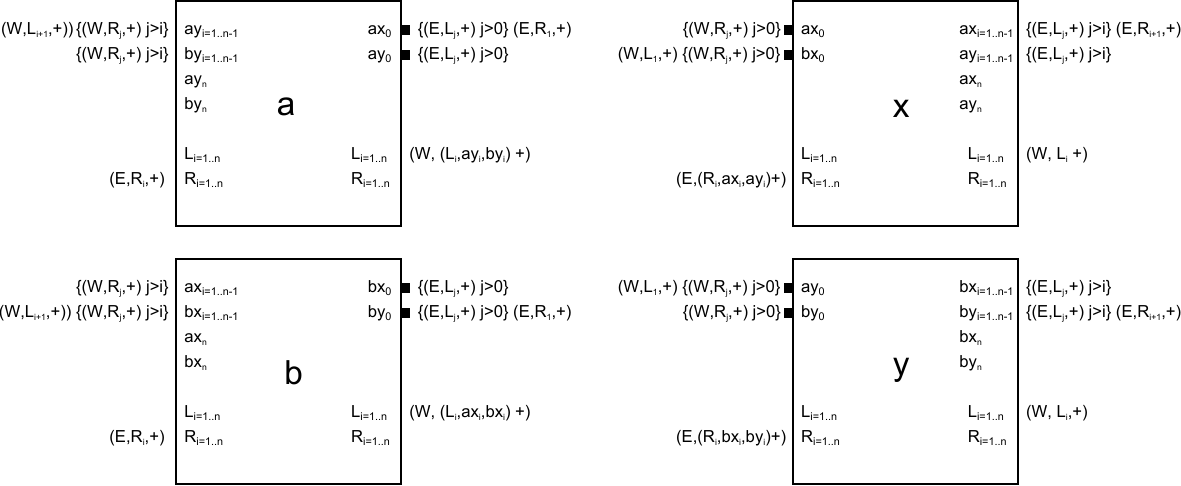}
\caption{\scriptsize The tileset for producing a $1 \times n$ line, where $n$ is a power of $2$.}
\label{fig:line_tiles}
\end{center}
\end{figure}

\begin{figure}
\begin{center}
\includegraphics[width=0.98\linewidth]{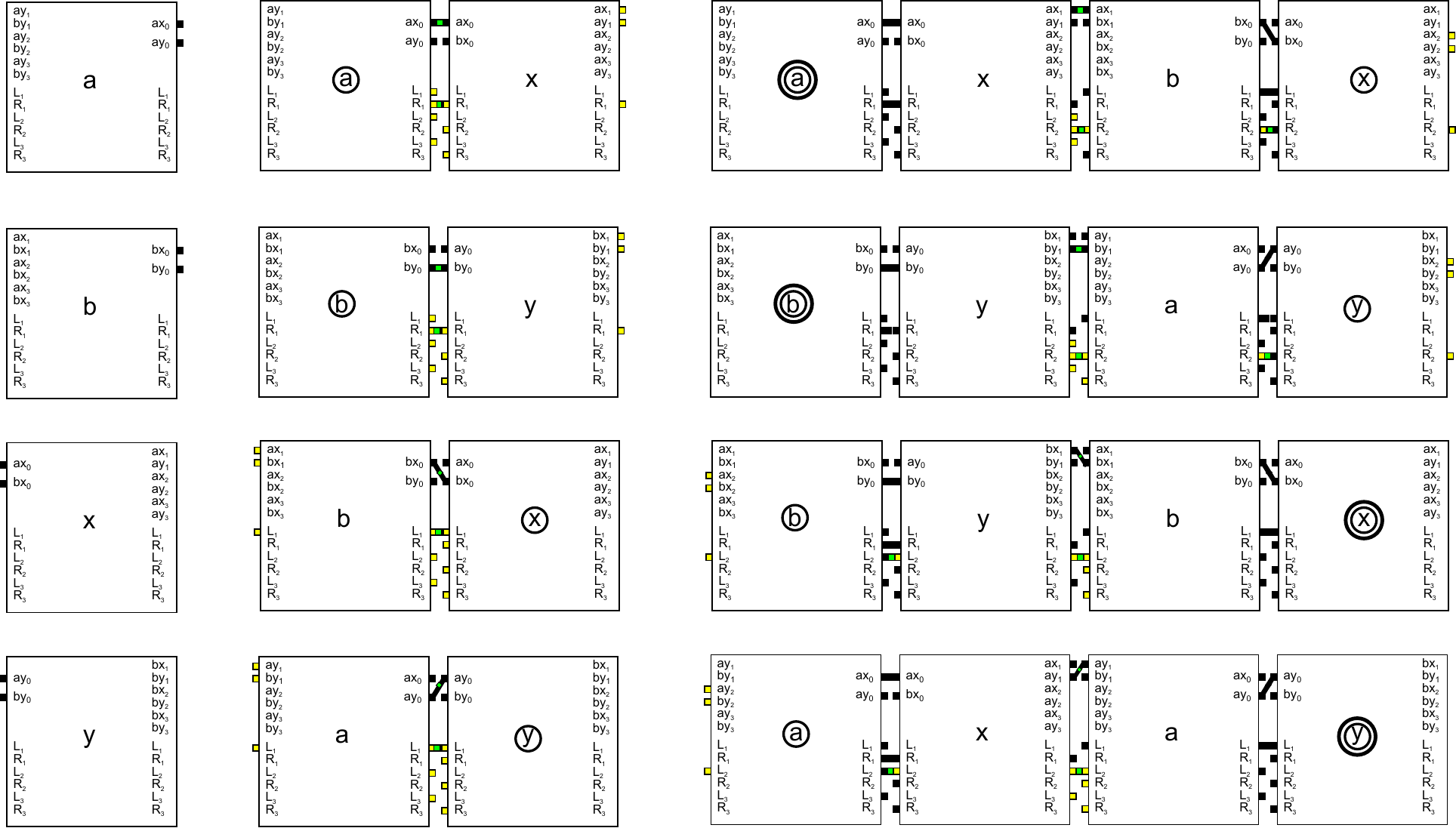}
\caption{\scriptsize Example assembly of a line of length $2^4 = 16$, or $n=4$  (part 1 of 2).
Circles indicate the identity of a block of tiles at a particular stage given by the number of circles.
For example, in the upper right, $a$ is circled twice to indicate that this block of four tiles has
identity $a_2$.}
\label{fig:exampleLine1}
\end{center}
\end{figure}

\begin{figure}
\begin{center}
\includegraphics[width=0.98\linewidth]{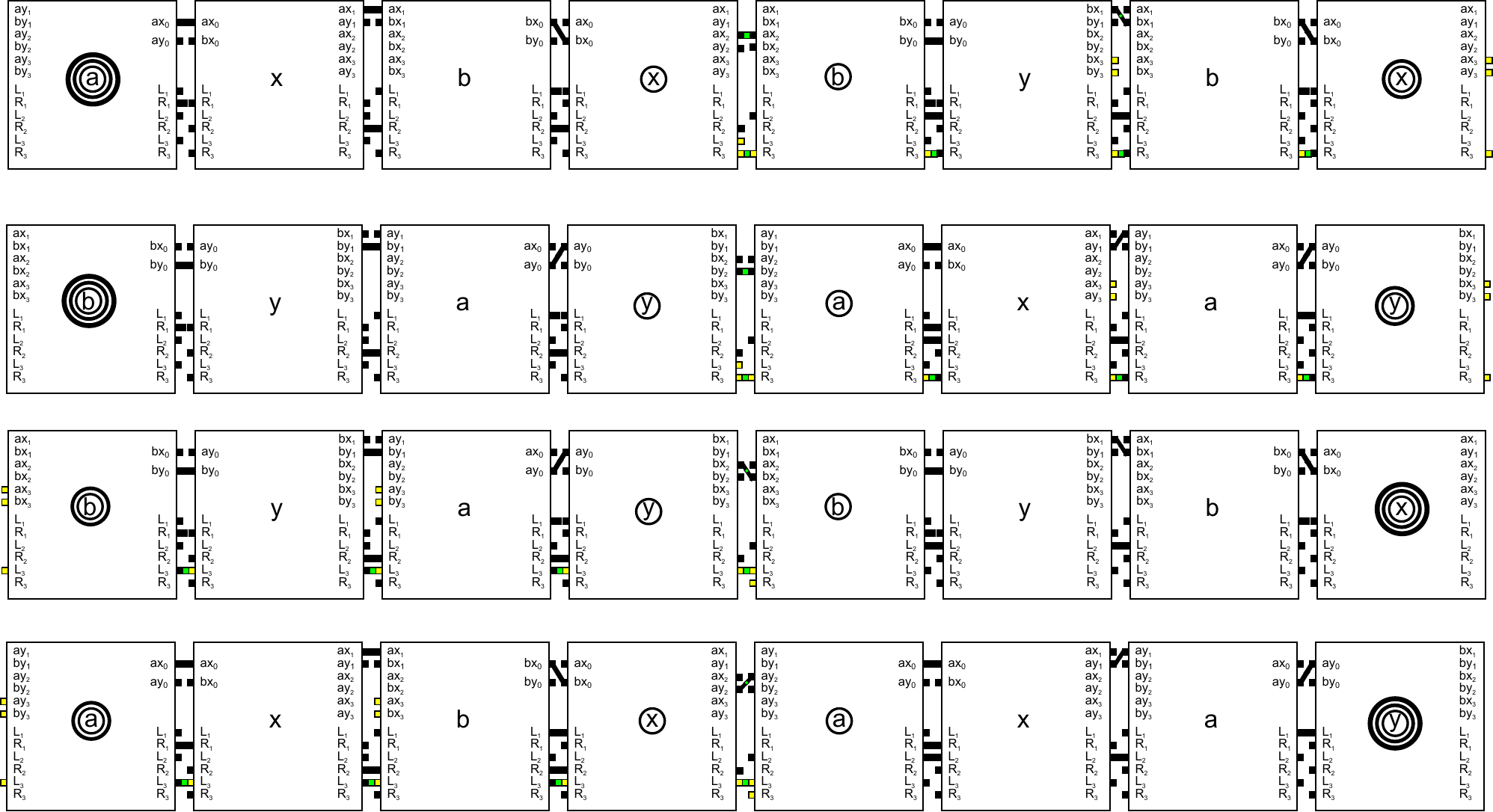}
\caption{\scriptsize Example assembly of a line of length $2^4 = 16$, or $n=4$  (part 2 of 2).  The final step of combinations, where each length $8$ segment combines with another to form a terminal assembly of length $16$ and which cause no glue activations, is not shown.}
\label{fig:exampleLine2}
\end{center}
\end{figure}

\end{proof}

\subsection{Proof of Theorem~\ref{thm:nline}}

\begin{theorem}\label{thm:nline}
For every $n \in \mathbb{N}$, there exists an STAM system $\mathcal{T} = (T,1)$, with $|T| = O(1)$, which uniquely assembles an $n \times 1$ line. Moreover, the signal complexity of $T$ is $O(\log{n})$ and $\mathcal{T}$ does not use glue deactivation.
\end{theorem}

\begin{proof}

Let $n \in \mathbb{N}$ be an arbitrary positive integer.  Note that $n$ can be written as a sum of powers of $2$, i.e. $n = \sum^{i=0}_{\lfloor \log n \rfloor} 2^i b_i$ where $b_i$ is the $i$th bit of the binary representation of $n$.
Just as $n$ can be written this way, the line of length $n$ can be composed of
segments with power of $2$ lengths, each composed during the construction of the line of length $2^k$ where $k = \lfloor \log n \rfloor$, as
described in the previous section. The joining of line segments corresponding to the powers of $2$ that sum to $n$ is shown in Figure~\ref{fig:Nline_scheme}.

 \begin{figure}
\begin{center}
\includegraphics[width=0.98\linewidth]{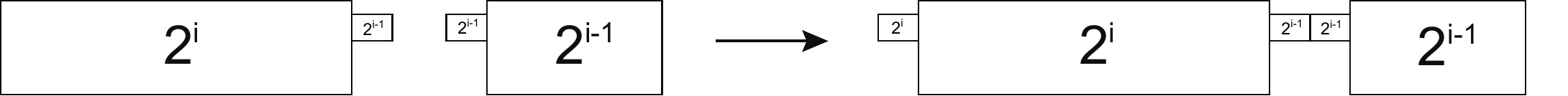}
\caption{Line segments of lengths corresponding to the powers of 2 that sum to $n$ must be joined
 to produce line of length $n$. Once line segments of unequal length have joined, signals ensure that growth
 continues only by addition of the correct unequal line segments. }
\label{fig:Nline_scheme}
\end{center}
\end{figure}

The tiles from the previous section are modified to include $O(\log n)$ more glues each
that direct the assembly of the powers of $2$ shown in Figure~\ref{fig:Nline_scheme}. The resulting tile set still has signal complexity $O(\log n)$ and is shown in Figures~\ref{fig:Nline_tiles1}-\ref{fig:Nline_tiles2}.  Depending on the specific bit sequence of $n$, the tile types $a'$, $x'$, and $x''$ may or may not be included in the tile set for that $n$.  For each power of $2$ which is a component of the binary representation of $n$, there is an assembly which stops doubling at that length and exposes a $j_i$ glue on its eastern edge that allows it to bind to the bar of length corresponding to the next lower power of $2$.  If $n$ is odd, then the $x_0$ tile is the only tile with an initially exposed $j$ glue (namely, $j_0$) on its west side.  If $n$ is even and $b_1 = 1$ (the second bit in its binary representation is $1$), then tile $a'$ is used in the construction and will present the first west facing $j$ glue, namely $j_1$.  If neither $b_0$ or $b_1$ are $1$, then tile type $x''$ is used to ensure that there will be lines of length equal to the smallest included power of two which will expose a $j$ glue on the west and will stop any continued doubling.  For all other included powers of $2$ (assuming that $n$ is not a power of $2$), the $x'$ tile will be included in the tile set and will ensure that bars of lengths corresponding to each such power of $2$ will stop doubling at those lengths and present $j$ glues on their east side.

Once a bar binds via a $j$ glue on its east, it sends a signal to the west side to open the next $j$ glue and allow subsequent attachment.  Finally, the bar of the largest power of $2$ will also stop doubling and send a signal (starting from either an $a$ or $b$ tile) to the east allowing for the attachment of the line consisting of all smaller powers of $2$.

 \begin{figure}
\begin{center}
\includegraphics[height=6.0in]{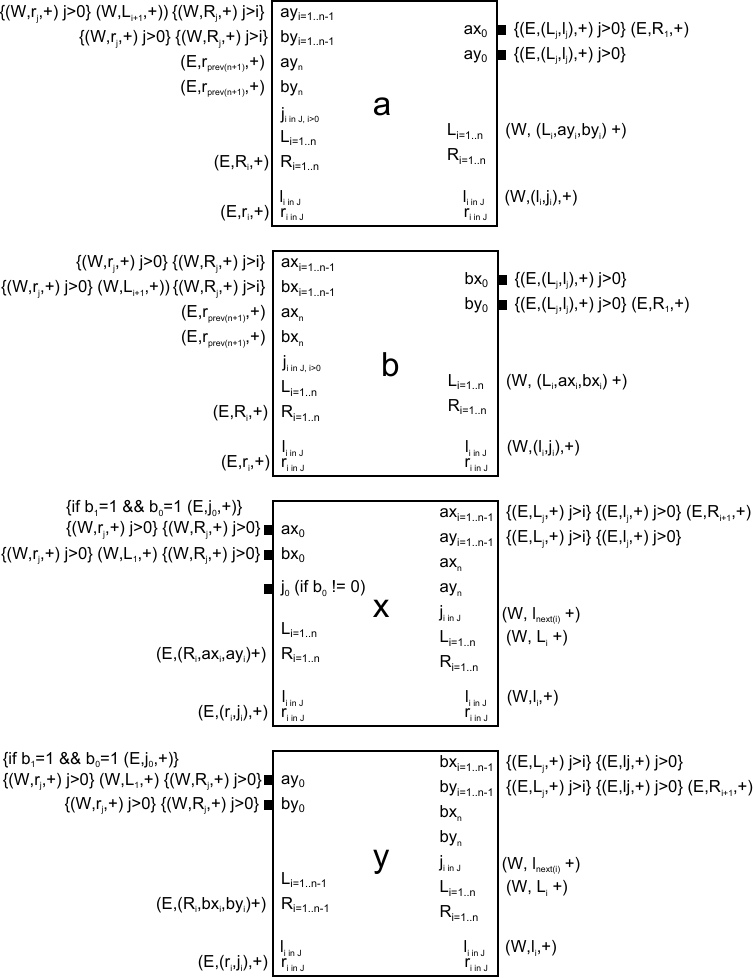}
\caption{The tile set (part 1 of 2) for self-assembling an $n \times 1$ line for arbitrary $n$. Notation:  $b = bin(k)$, i.e. $k$ represented in binary, e.g. for $k = 19$, $bin(k) = 10011$.  $b_i$ = $i$th bit of $b$ (where $b_0$ is the least significant bit), e.g. for $k=19$, $b_0=1$, $b_1=1$, $b_2=0$, $b_3=0$, and $b_4=1$. $n = \lfloor \log k \rfloor$, e.g. for $k = 19$, $n = 4$. $I = \{ i | b_i = 1\}$ as an ordered set, e.g. for $k = 19$, $bin(k) = 10011$, $I = \{0,1,4\}$. $J = I - max(I)$, namely $I$ without the largest element, e.g. if $I = \{0,1,4\}$ then $J = \{0,1\}$. $prev(i) =$ the element of $J$ immediately less than $i$, e.g. $prev(4) = 1$. $next(i) =$ the element of $J$ immediately greater than $i$, e.g. $next(0) = 1$.}
\label{fig:Nline_tiles1}
\end{center}
\end{figure}

 \begin{figure}
\begin{center}
\includegraphics[height=5.0in]{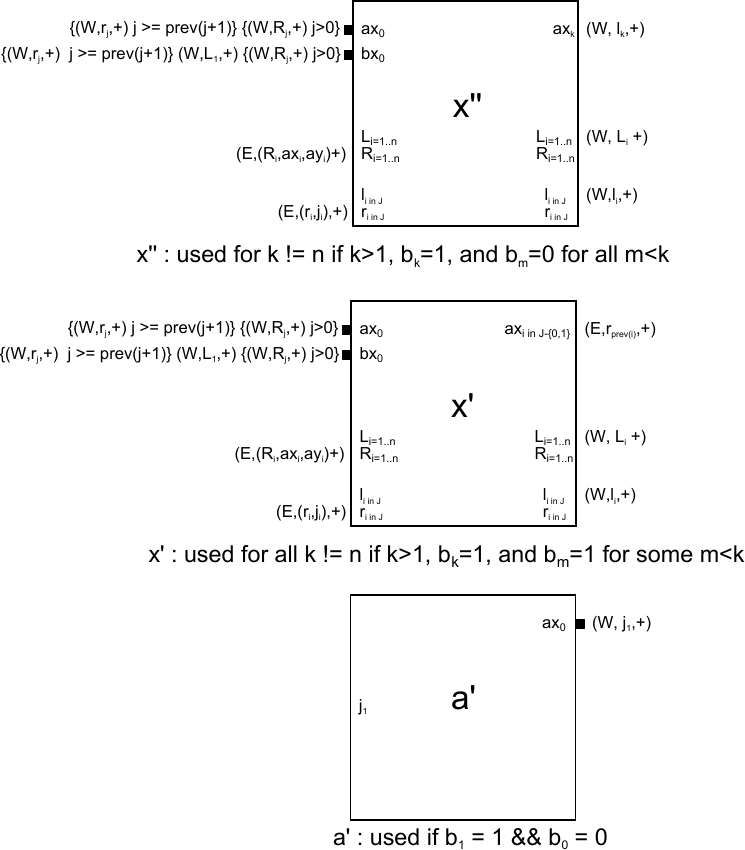}
\caption{The tile set (part 2 of 2) for self-assembling an $n \times 1$ line for arbitrary $n$. See Figure~\ref{fig:Nline_tiles1} for notation.}
\label{fig:Nline_tiles2}
\end{center}
\end{figure}

\begin{figure}
\begin{center}
\includegraphics[width=0.98\linewidth]{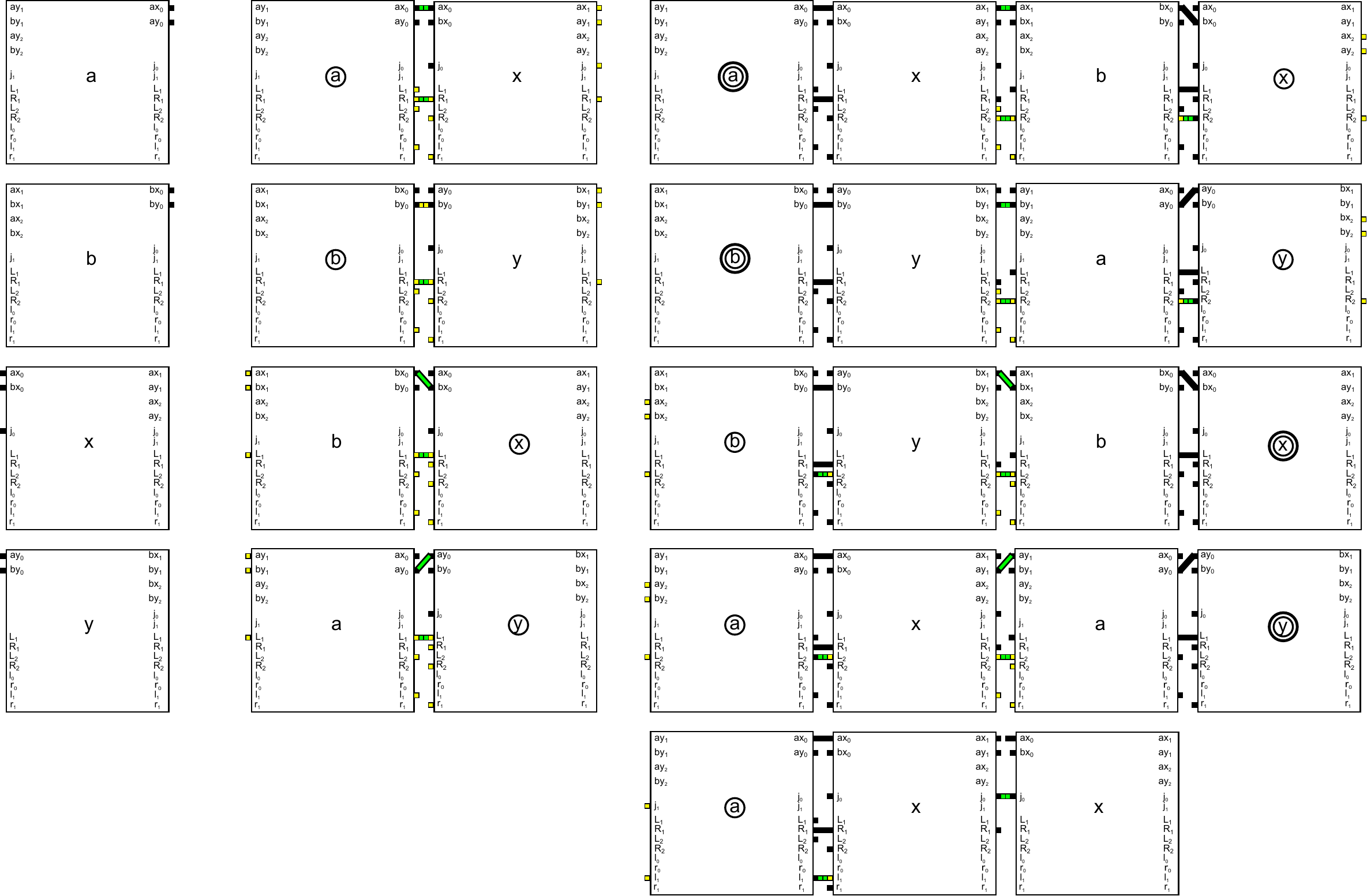}
\caption{\scriptsize Example assembly of a line of length $11$ (part 1 of 2).}
\label{fig:Nline_example1}
\end{center}
\end{figure}

\begin{figure}
\begin{center}
\includegraphics[width=0.98\linewidth]{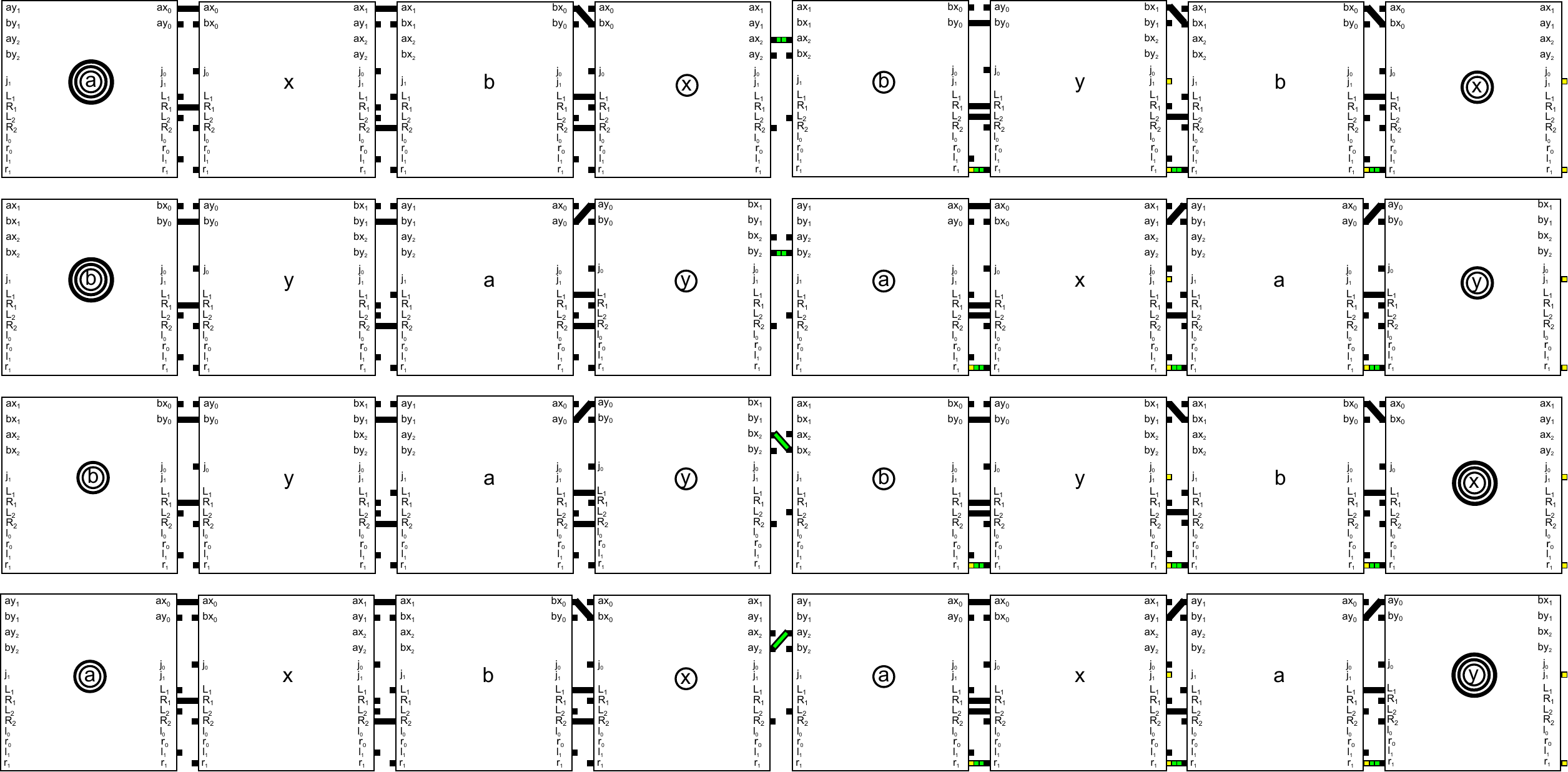}
\caption{\scriptsize Example assembly of a line of length $11$ (part 2 of 2).  The final step of combinations, where each length $8$ segment combines with the single assembly of length $3$ from Figure~\ref{fig:Nline_example1} is not shown since there are no additional glue activations which occur, merely the binding of each assembly of length $8$ with one of length $3$ to form final line assemblies of the target length, $11$.}
\label{fig:Nline_example2}
\end{center}
\end{figure}

\end{proof}

\section{Fuel Efficient Turing Machines}\label{sec:TM}

Showing that the original abstract Tile Assembly Model is computationally universal is a simple matter of designing a tile assembly system which can simulate a universal Turing machine, as originally shown in \cite{Winf98}, and later expanded upon in \cite{jSADS,RNAPods,CookFuSch11,DotKarMasNegativeJournal}.  While displaying the computational power of the aTAM (and variants prior to the STAM), a common drawback of the constructions has been the number of tiles utilized during the formation of the assembly which simulates the computation, which, in this paper, is referred to as the fuel efficiency of the simulation.

For prior constructions, it has been necessary to make a new copy of the entire tape of the Turing machine between each computational step, with the new copy identical to the original except for the slight difference of a mere two tape cells indicating: 1. the output value in the tape cell left by the tape head, and 2. the tape cell marking the current location of the tape head.  This full-scale copying of the tape, including the vast majority of cells which are unchanged, is wasteful in terms of the number of tiles required, experimentally very error prone due to the huge number of tile attachments required, and results in enormous assemblies.
In this section, we exhibit a construction which is capable of simulating a universal Turing machine in the STAM, but while doing so only requires a small constant number of tiles (never more than $7$) as fuel for each computational step and maintains an assembly which consists of a number of tiles which is only twice the total number of tape cells used by the Turing machine up to that step.  It is possible that with significantly fewer binding events in the STAM construction than in those of previous models (even taking into account those used for signaling), it may be the case that the number of errors which occur could decrease, assuming, of course, that the mechanism which carries out the transition function is sufficiently error-free.

Throughout this paper, and without loss of generality, we define Turing machines as follows.
Let $M$ be an arbitrary single-tape Turing machine, such that $M = (Q,\Sigma,\Gamma,\delta,q_0,q_{\textmd{accept}},$ $q_{\textmd{reject}})$ with state set $Q$, input alphabet $\Sigma = \{0,1\}$, tape alphabet $\Gamma = \{0,1,\_\}$, transition function $\delta$, start state $q_0 \in Q$, accept state $q_{accept} \in Q$, and reject state $q_{reject} \in Q$.  Furthermore, $M$ begins in state $q_0$ on the leftmost cell of the tape, expects a one-way infinite-to-the-right tape, and is guaranteed to never attempt to move left while on the leftmost tape cell.

\begin{theorem}[Fuel efficient Turing machines]
    \label{thm:TM}
    For any Turing machine $M$ with input $w \in \{0,1\}^*$, there exists an STAM system $\mathcal{T}_{M(w)} = (T_{M(w)}, 1)$ with tile complexity $O(|Q|)$, signal complexity $O(1)$, and fuel efficiency $O(1)$, which simulates $M$ on $w$ in the following way:
    \begin{enumerate}
        \item $T_{M(w)}$ contains an active supertile consisting of $2|w|+2$ active tiles representing $w$ and $M$'s start state.

        \item If $M$ halts on $w$, then $\mathcal{A}_{\Box}[T_M,1]$ contains exactly one supertile with $> 3$ tiles and that supertile contains exactly one $ACCEPT$ ($REJECT$) tile if $M$ accepts (rejects) $w$.

        \item If $M$ does not halt on $w$, then $\mathcal{A}_{\Box}[T_M,1]$ contains exactly $0$ (terminal) supertiles with $> 3$ tiles.

    \end{enumerate}
\end{theorem}

Our proof of Theorem~\ref{thm:TM} is by construction.
Here, we provide a brief overview.

Our construction works by utilizing a set of tile type templates that, along with the definition of a Turing machine $M$, are used to generate the set of active tiles which are specific to $M$.  The construction uses a pair of tiles to represent each tape cell, with one tile representing the value ($0$, $1$, or $\_$) of that cell and one tile providing a ``backbone'' which the other attaches to and which also attaches to the backbone tiles of the cells to its left and right.  Additionally, there is a special tile for the tape cell representing the rightmost end of the tape, and also, at any given time, exactly one tape cell which also represents one state of $M$ along with the tape cell value.  The location of the tape cell with that information denotes the location of $M$'s tape head at that point, and the value of the state tells what state $M$ is in.  Transitions of $M$ occur in a series of $4$ main steps in which tiles bind to the north of the tape cell denoting the head location, then to the north of the tape cell to the immediate left or right (depending on whether or not $M$'s transition function specifies a left or right moving transition from the current state while reading the current tape cell value), and along the way cause the dissociation of the tiles representing the tape cell values in both locations and their replacement with tiles which represent the correct output tape cell value of the transition and correctly record the new state and head location at the tape cell immediately to the left or right.  Due to the asynchronous nature of glue deactivations, and also the necessity that any ``junk'' assemblies produced (i.e. those assemblies which break off from the assembly representing the Turing machine tape and which don't contribute to the final ``answer'') must not be able to attach to any portion of any supertile which represents any stage of the computation, junk assemblies are produced as size $2$ or $3$ so that any activated glues which would otherwise be able to bind to another supertile are hidden between the tiles composing the junk assembly.  In such a way, $M(w)$ is correctly simulated while requiring only a constant number of new tiles per simulated transition step, and all junk assemblies remain inert and at size either $2$ or $3$.  If $M(w)$ halts, there will be one unique, terminal supertile which represents the result of that computation and is of size $> 3$.  If $M(w)$ does not halt, only the junk assemblies will be terminal.

\begin{proof}[Proof sketch]

\begin{figure}[htp]
\begin{center}
\includegraphics[width=5.5in]{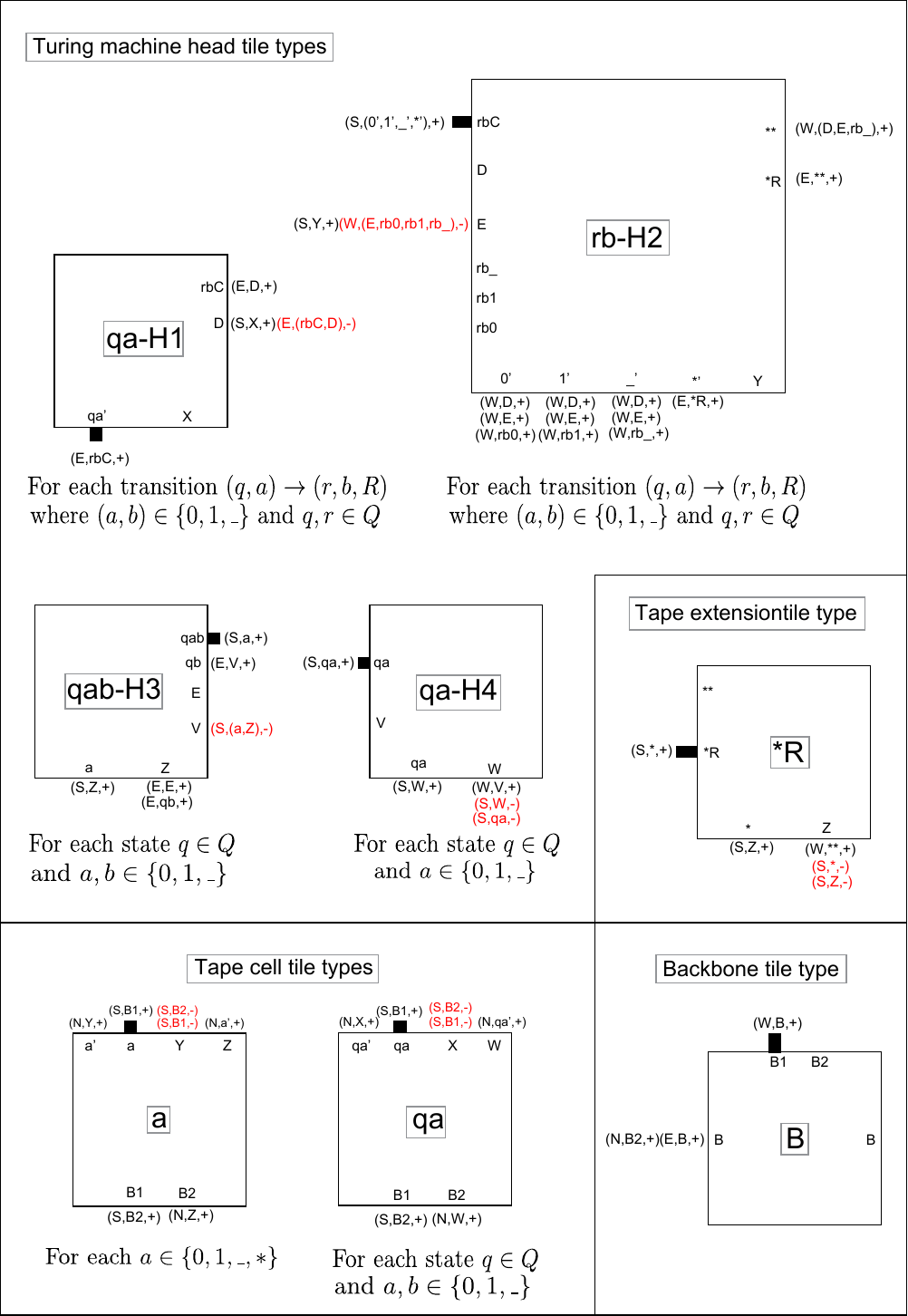}
\caption{
The templates used to generate tile types for the simulation of a given Turing machine $M$ with state set $Q$, tape alphabet $\{0,1,\_\}$, and the specified transitions which move the head right.  The necessary tile types to simulate transitions which move the head to the left and simply mirror images of the ``Turing machine head tile types''.}
\label{fig:TM-tile-designs}
\end{center}
\end{figure}

Our proof is by construction. Let $M=(Q, \Sigma, \Gamma, \delta, q_0, q_{\textmd{accept}}, q_{\textmd{reject}})$ be a Turing machine and $w \in \{0,1\}^*$ be the input. We define the finite set of tile types $T_{M(w)}$ in relation to $M$ as shown in Figure \ref{fig:TM-tile-designs}.  Note that the tile types presented are generic templates applicable for generating the tile types specific to a given $M$ (with the exception of the ``tape extension tile type'', ``backbone tile type'', and the left ``tape cell tile type'' which are used for all $M$).  Also note that the tile type templates provided for the ``Turing machine head tile types'' are specific to transitions which move the tape head to the right, and those for left-moving transitions are simply mirror images with a unique set of glues (e.g. every glue is prefixed with ``$L$-'') defined specifically for those transitions.  The actual tile types that would be generated for a specific $M$ can be determined by substituting each valid variable value (as specified below each tile type template) into glue labels, and potentially consist of a large number of actual tile types generated by each template definition.

The tiles in the ``Tape cell tile types'' and ``Backbone tile type'' sections of Figure~\ref{fig:TM-tile-designs} are used to compose the tape. The tape itself, as shown in Figure~\ref{fig:TMFigureExample_Seed}, consists of a row of east-west connected ``backbone'' tiles, each of which is connected on its north to a tile representing a tape cell.  The backbone is used to keep the entire assembly connected during the process of tape cell replacement.  The possible values for tape cells are: (1) ``$0$'', (2) ``$1$'', (3) ``\_'' (a blank), (4) ``*'' (which represents the currently rightmost tape cell), or (5) $q \times \{0,1,\_\}$ where $q \in Q$ (at any point there is exactly one such tape cell which represents the location of the head and $M$'s current state along with the contents of the tape cell that the head is currently reading.  The tiles in the ``Turing machine head tile types'' group are used to simulate the actions of the head, which can be understood as four high-level steps ($H1$,$H2$,$H3$, and $H4$ - to be explained shortly). The ``Tape extension tile type'' is used to extend the tape by one additional blank symbol (``\_'') if and when the head reaches the rightmost end of the tape, thus allowing the assembly to simulate a one-way infinite-to-the-right tape.

\begin{figure}[htp]
\begin{center}
\includegraphics[width=3.0in]{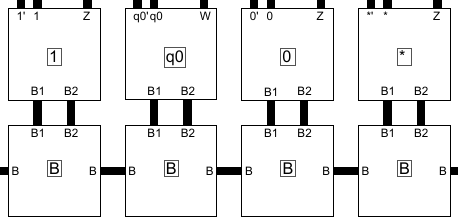}
\caption{
Start state of the Turing machine $M$ simulation example assuming a tape of ``$100*$'' with $M$ in state $q$ and currently reading the leftmost $0$.}
\label{fig:TMFigureExample_Seed}
\end{center}
\end{figure}

The easiest method of explanation of this construction is to step through an example transition, which we now do.  The specific example is depicted in the series of Figures~\ref{fig:TMFigureExample_Seed}, \ref{fig:TM-high-level-example1} and \ref{fig:TM-high-level-example2}. Assume that $M$ is currently in state ``$q$'' and reading symbol ``$0$'', the output symbol is ``$1$'', the next state should be ``$r$'', and the head should move right (i.e. the simulated transition is $(q,0) \rightarrow (r,1,R)$). Glues which are currently \texttt{on} are shown as black, glues which have just been turned \texttt{on} or connected in the current step are shown as bright blue, glues which are queued to turn \texttt{off} are shown as red, and glues which are currently \texttt{off} or \texttt{latent} have been hidden. Triggered actions have been depicted by arrows.

\begin{figure}[htp]
\begin{center}
\includegraphics[width=6.0in]{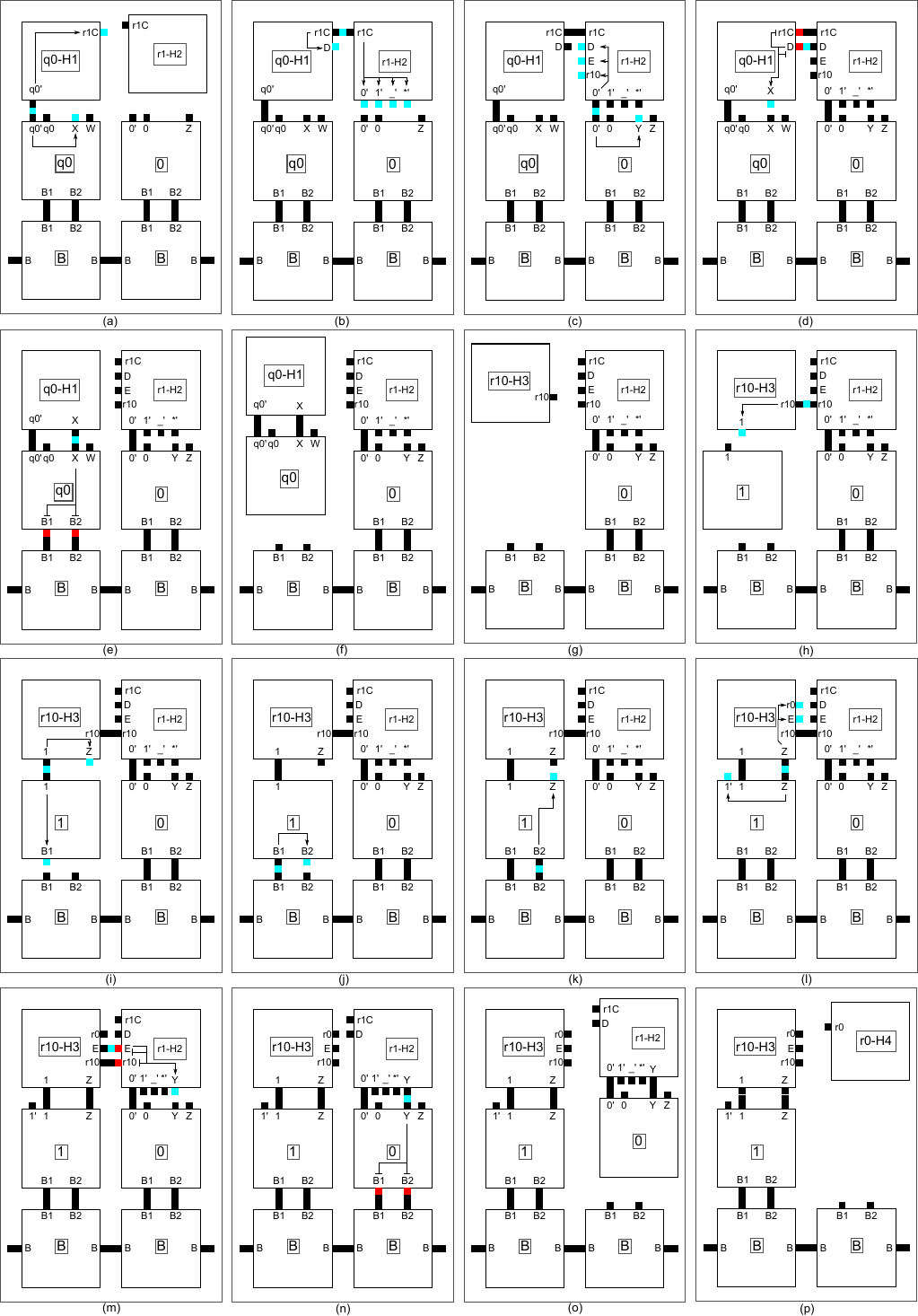}
\caption{
High-level sketch of the simulation of a transition, part 1 of 2}
\label{fig:TM-high-level-example1}
\end{center}
\end{figure}

\begin{figure}[htp]
\begin{center}
\includegraphics[width=6.0in]{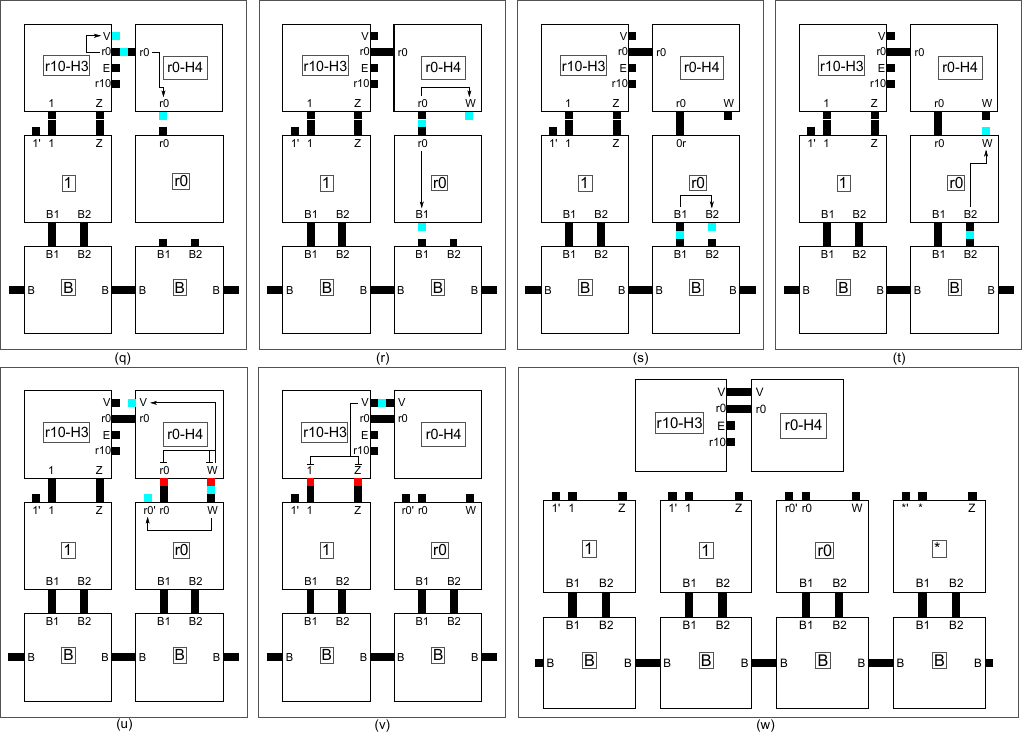}
\caption{
High-level sketch of the simulation of a transition, part 2 of 2}
\label{fig:TM-high-level-example2}
\end{center}
\end{figure}

The transition begins from the $q0'$ glue on the $q0$ tile. As depicted in Figure~\ref{fig:TM-high-level-example1}(a), the $q0$-$H1$ tile encodes the information to read a $0$ while in state $q$, change it to a $1$, move the head to the right, and change to state $r$. The values $r$, $1$ and ``move to the right'' are implicitly stored by $r1$-$H2$ as shown in Figure~\ref{fig:TM-high-level-example1}(b). $r1$-$H2$ then connects to the tape cell tile below to read the symbol on that cell (a $0$ here), and sends a message back to $q0$-$H1$ which causes $q0$-$H1$ and the now obsolete tape cell beneath it to dissociate as a pair, as shown in Figure~\ref{fig:TM-high-level-example1}(c)-Figure~\ref{fig:TM-high-level-example1}(f). When $r1$-$H2$ connected with $q0$-$H1$, $r1$-$H2$ didn't know what the tape cell value beneath it was, so it had to activate all four possible glues, which are $0'$, $1'$, $\_'$, and $*'$. Since only the $0'$ binds, $r1$-$H2$ can pass along the value of $0$ in its next glue $r10$ to a tile of type $r10$-$H3$. Here the $r10$ means the next state will be $r$, the tape cell of the current head position will change to $1$, and the symbol of next head position is $0$. Then, in Figure~\ref{fig:TM-high-level-example1}(h)-Figure~\ref{fig:TM-high-level-example1}(l), the $r10$-$H3$ tile causes a new tile representing a tape cell value of $1$ to fully bind to the backbone and activate its $1'$ glue, which completes the transition of that tape cell. After that $1$ tile is connected to the backbone, it sends a message to $r1$-$H2$ that allows it to dissociate, as shown in Figure~\ref{fig:TM-high-level-example1}(l)-Figure~\ref{fig:TM-high-level-example1}(o). Finally, the $r0$-$H4$ tile binds to $r10$-$H3$'s $r0$ glue, allowing it to facilitate the binding of the $r0$ where the $0$ previously was, and eventually dissociate along with the $r10$-$H3$ once the $r0$ is fully connected, as shown in Figure~\ref{fig:TM-high-level-example1}(p)-Figure\ref{fig:TM-high-level-example2}(w).  At this point, the tape is ready for next state transition, in a configuration similar to that at the beginning of the transition but with the correct output from the transition which is now complete, and the head in the correct position while representing the new state.

\subsection{Growing the tape}

In order to simulate a one-way infinite-to-the-right tape with a finite assembly, we simply design the tape so that the rightmost cell always represents a special value, namely $*$.  Whenever a transition begins which needs to move the tape head to the right, and the destination location of the head currently contains the $*$ symbol, that situation is ``read'' by the $rb$-$H2$ tile which forces a $*R$ tile to bind to its right, which itself ensures that a new tape cell location with the $*$ symbol as well as a new backbone tile to hold it in place at the end of the tape both firmly bind.  Additionally, the $rb$-$H2$ tile ``fakes'' the situation of having read a blank ($\_$) symbol so that the location previously occupied by the $*$ is treated as though it was occupied by a $\_$, which - coupled with the new $*$ symbol one location to the right - is logically identical to adding a blank tape cell on the right end of the tape.  Since this occurs whenever the head tries to move to the end of the tape, it simulates an infinite tape.

\subsection{Junk assemblies}

\begin{figure}[htp]
\begin{center}
\includegraphics[width=3.0in]{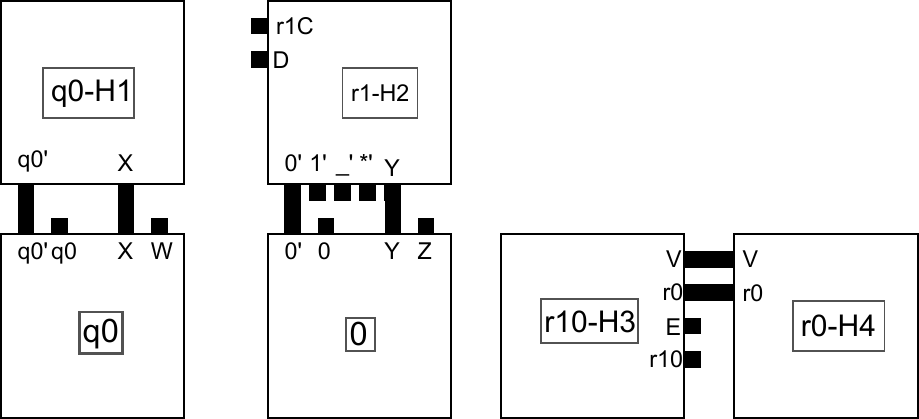}
\caption{
``Junk'' assemblies produced during the example state transition shown in Figures~\ref{fig:TM-high-level-example1} and \ref{fig:TM-high-level-example2}}
\label{fig:TMFigureExample_Junk}
\end{center}
\end{figure}

During each transition of $M$, several junk assemblies are created when they dissociate from the main assembly representing the tape.  These can be seen in Figure~\ref{fig:TM-high-level-example1}(f), Figure~\ref{fig:TM-high-level-example1}(o), and Figure~\ref{fig:TM-high-level-example2}(w),  and are also shown in Figure~\ref{fig:TMFigureExample_Junk}.  In order to guarantee correctness of the construction and thus the underlying simulation of $M(w)$, it is necessary that these junk assemblies cannot interfere with \emph{any version} of the computation (keeping in mind that many such simulations would be occurring in parallel).  To ensure this, we have carefully designed the junk assemblies so that every glue on their perimeter must be either \texttt{latent} or \texttt{off}, with the notable exception of the glues $r1C$ and $D$ on the west side of the $r1$-$H2$ tile of the middle junk assembly shown in Figure~\ref{fig:TMFigureExample_Junk}.  The method of ensuring that specific glues are actually deactivated in a temperature $\tau=1$ system is to (1) only send deactivation signals to glues which must be bound at the time the signal is sent and (2) to only send deactivation signals to glues on one side of a bond (i.e. if two tiles are bound together by glue $x$, only one of those tiles sends a signal to deactivate its copy of $x$).  In this way, despite the fact that signals are processed asynchronously, it must be the case that if a supertile $\gamma$ is going to detach, all glues on the boundary of $\gamma$ (whether on $\gamma$ or the supertile that $\gamma$ is bound to) which have been sent a deactivation signal have actually already turned that glue \texttt{off}.  Furthermore, since it is often the case that sets of glues must be turned \texttt{on} with the expectation that only one of them will ever bind (e.g. as in the case when the $rb$-$H2$ tile activates one glue for each potential tape cell value to its south), in order to follow (1) above, it is necessary to conceal the unused \texttt{on} glues from any potential interactions.  This need to segregate such glues is the reason for creating junk assemblies of size $2$ (and $3$) rather than of size $1$, and is a technique utilized in all of the constructions which make use of glue deactivation within this paper.

Following the above design techniques makes it necessary that for the leftmost and rightmost junk assemblies to detach, they must be completely inert and unable to bind to any other supertiles in the system.  The middle supertile, with the $r1C$ and $D$ glues still active, on the other hand, ensures that it cannot bind to any other supertile by utilizing a form of geometric hindrance.  Essentially, the $r1$-$H2$ tile can only ever dissociate if it does so with a tape cell tile attached to its south.  While the $r1C$ glue on its west would otherwise be able to attach to a $qa$-$H1$ tile that is currently bound to the tape using the eastern $rbC$ (in this case $r1C$) glue, an $H1$ tile can only ever attach to a tape cell tile after an $H3$ tile dissociates from its north.  The dissociation of the $H3$ can only happen if it does so while connected to an $H4$ tile, and the $H4$ tile could only dissociate along with the $H3$ tile after ``filling in'' the new tape cell tile.  Thus, for the western facing $rbC$ glue on an $H1$ tile to be available for binding, it can never be the case that a junk supertile like that pictured in the middle of Figure~\ref{fig:TMFigureExample_Junk} would be able to bind:  it is too tall (the necessary space for the southern tile would already be occupied by a tape cell tile).

Finally, in the case where the tape is grown one cell to the right there is a fourth type of junk assembly.  This junk assembly is simply the middle junk assembly of Figure~\ref{fig:TMFigureExample_Junk} with a $*R$ tile bound to the east of the $rb$-$H2$ tile by both $**$ and $*R$ glues (for a total size of $3$ tiles).  However, the southern glues of the $*R$ tile must be deactivated before dissociation of the junk assembly, and therefore no new glues in the \texttt{on} state are added.  As this is the only other producible type of junk assembly, this means that all junk assemblies are fully inert and unable to bind to any other supertiles, and thus maintain the correctness of the simulation while also never growing larger than size $3$.

\subsection{Correctness of Theorem~\ref{thm:TM}}

It has been shown that the STAM system $\mathcal{T}_{M(w)}$ which simulates $M(w)$ correctly simulates the behavior of TM $M$ on input $w$.  If $M(w)$ halts, special $qa$-$H1$ tiles for $q \in \{q_{accept},q_{reject}\}$ (previously undiscussed) will attach in the head location which halt further assembly (by containing no other \texttt{on} or activatable glues), thus creating a terminal assembly of size $> 3$ (since there must be at least one tape cell tile representing a $0$, $1$, or $\_$ and the tape cell tile representing $*$, along with their backbone tiles).  If $M(w)$ does not halt, the supertile representing the tape will never be terminal - only the junk assemblies will be terminal - and thus there will be no terminal supertiles of size $> 3$.  Furthermore, it is a temperature $\tau=1$ system, all junk assemblies remain at size either $2$ or $3$, the signal complexity is $O(1)$ (specifically, $6$ due to the maximum number of glues on a tile side being the $6$ on the west side of $rb$-$H2$), the fuel efficiency is $O(1)$ as every transition produces exactly $3$ junk assemblies, each with size $\leq 3$, and adds at most $4$ new tiles to the tape assembly (the two new tiles for the tape cells which have swapped the head position, plus possibly two more tiles if the tape was grown by one symbol to the right), and tile complexity $O(|Q|)$ since the tape alphabet and all other glue values are a constant size set and each tile template can be used to generate at most a constant number of tiles for each transition in $\delta$ and there can be at most $O(|Q|)$ transitions in $\delta$.
\end{proof}

\section{Self-Assembly of the Sierpinski Triangle}
\label{sec:triangle}

Discrete self-similar fractals are defined as sets of points in $\mathbb{Z}^2$, and consist of infinite, aperiodic patterns.  It is difficult, if not impossible, for them to strictly self-assemble in the aTAM, as is shown in \cite{jSSADST,jSADSSF} where the impossibility of a class of discrete self-similar fractals, including the Sierpinski triangle, strictly self-assembling in the aTAM is proven.  The impossibility of strictly self-assembling the Sierpinski triangle in the 2HAM was
shown in \cite{Versus}.
Additionally, Doty \cite{BoundedPinch} has shown a generalization of the impossibility proof from \cite{jSADSSF} which applies to, among other things, scaled versions of the Sierpinski triangle for any scaling factor.  Thus, any method of strictly self-assembling the Sierpinski triangle, scaled or not, is of interest.

In this section, we show that weak self-assembly of the Sierpinski triangle is possible in the STAM with fewer tile types (4 versus 7) and lower temperature (1 versus 2) than existing TAM constructions, and we also show that strict self-assembly at scale factor $2$ is possible in the STAM at temperature 1, a first for any model at any temperature.

\subsection{The discrete Sierpinski triangle} 
\begin{figure}[htp]
\vspace{-20pt}
\centering
  \subfloat[][The first five stages of the discrete Sierpinski triangle.]{%
        \label{fig:Sierpinski-triangle}%
        \centering
        \includegraphics[width=2.5in]{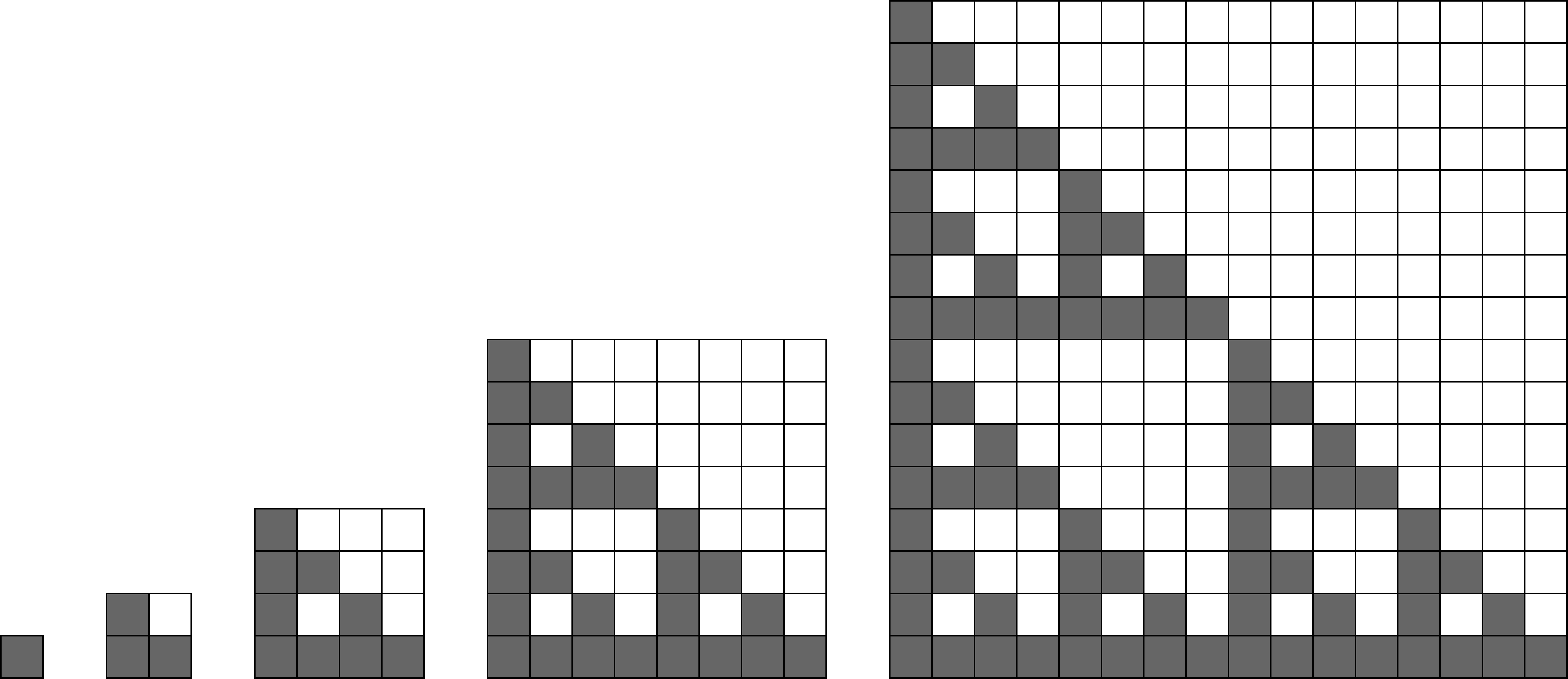}}%
        \hspace{15pt}
  \subfloat[][Transformation of a point $(x,y)$ into a set of points $f(x,y)$ for scale factor $2$, and associated notation.]{%
        \label{fig:ST-block}%
        \centering
        \includegraphics[width=2.0in]{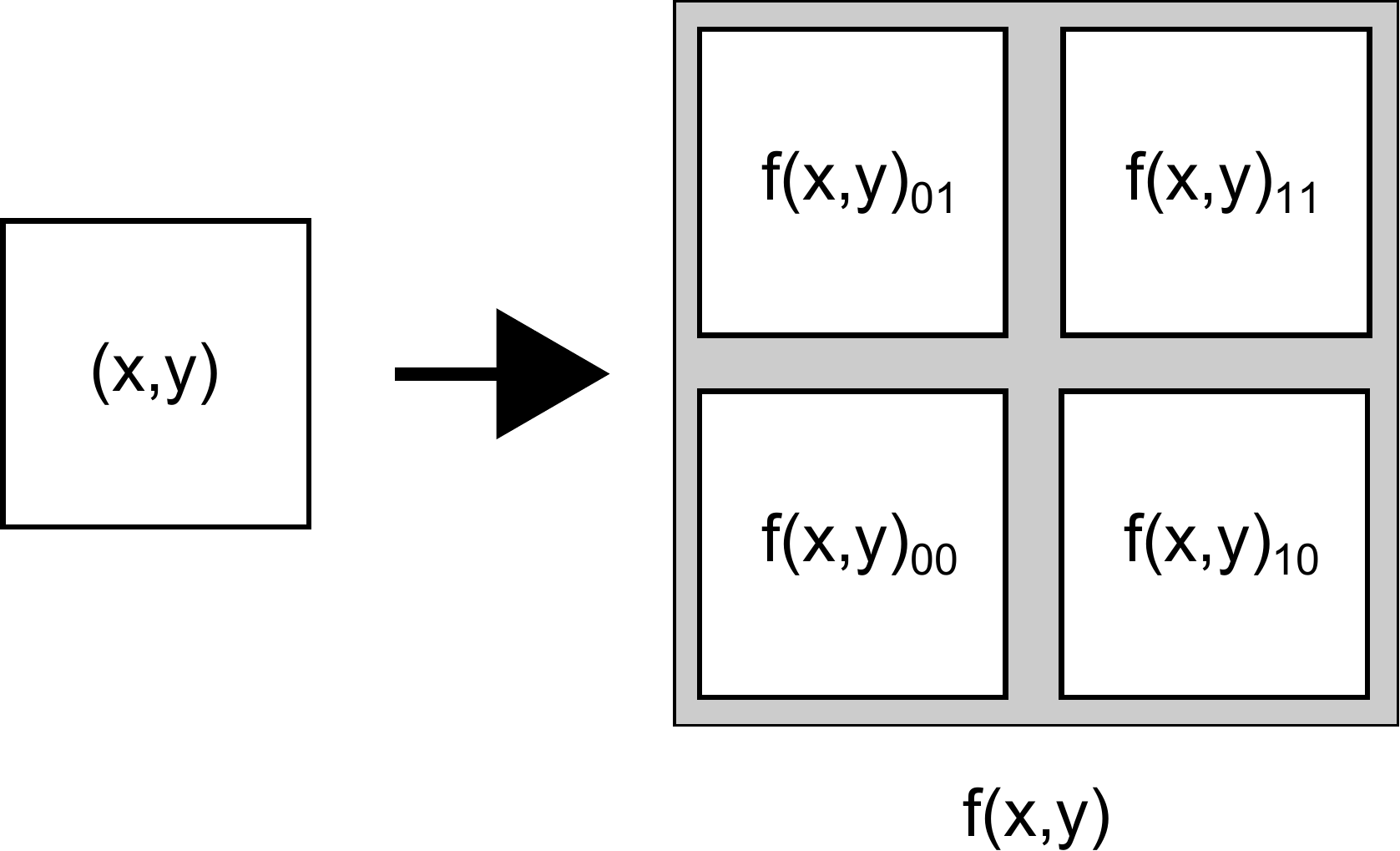}
        }%
  \centering
  \caption{The Sierpinski triangle and a description of the mapping for the scaled version.}
  \vspace{-20pt}
\end{figure}

Here we use the definition of \cite{jSSADST}.
Let $V = \{(1,0),(0,1)\}$.  Define the sets $S_0, S_1, S_2, \dots \subset \mathbb{Z}^2$ by the recursion $S_0 = \{(0,0)\}$, $S_{i+1} = S_i \cup (S_i + 2^i V)$, where $A + c B = \{\vec{m} + c \vec{n} \mid \vec{m} \in A$ and $\vec{n} \in B \}$.  Then the (standard) discrete Sierpinski triangle is the set $S_{\triangle} = \cup^{\infty}_{i=0}S_i$.  See Figure~\ref{fig:Sierpinski-triangle} for a depiction of the first five stages (i.e. $S_0$ through $S_4$).

Our Sierpinski triangle constructions are as stated in the following two theorems.
Additionally, in the next section we provide a high-level sketch of the more technically challenging construction for Theorem~\ref{thm:strict-Sierpinski}.

\begin{theorem}\label{thm:weakSierp}
There exists an STAM system that weakly self-assembles the Sierpinski triangle.  The system has 5 unique tiles, signal complexity $= 4$, assembles at temperature $\tau=1$, and does not utilize glue deactivation.
\end{theorem}

\begin{theorem}\label{thm:strict-Sierpinski}
There exists an STAM system that strictly self-assembles the discrete Sierpinski triangle at temperature $\tau=1$,
with tile complexity $= 19$, scale factor $= 2$, signal complexity $= 5$, and which makes use of glue deactivation, producing terminal junk assemblies of size $\leq 6$.
\end{theorem}

\subsection{Weak Self-Assembly of the Sierpinski Triangle}
\label{sec:weak-triangle-app}

Here we describe an STAM system that weakly self-assembles $S_{\triangle}$ to prove Theorem~\ref{thm:weakSierp} by construction.  Of note is that this construction works at temperature 1, thus using no cooperative binding, and does not utilize glue deactivation.  Further, the tile complexity and signal complexity are small constants of 5 tiles and 4 signals per tile face.  This construction is an example of a more general technique for simulating any $\tau=2$ rectilinear aTAM system (rectilinear systems grow from south to north, west to east) with a temperature $\tau=1$ STAM system.  The general simulation of any $\tau=2$ aTAM system with a temperature $\tau=1$ system STAM system is direction for future work.

\begin{figure}[here]
\begin{center}
\includegraphics[width=5.0in]{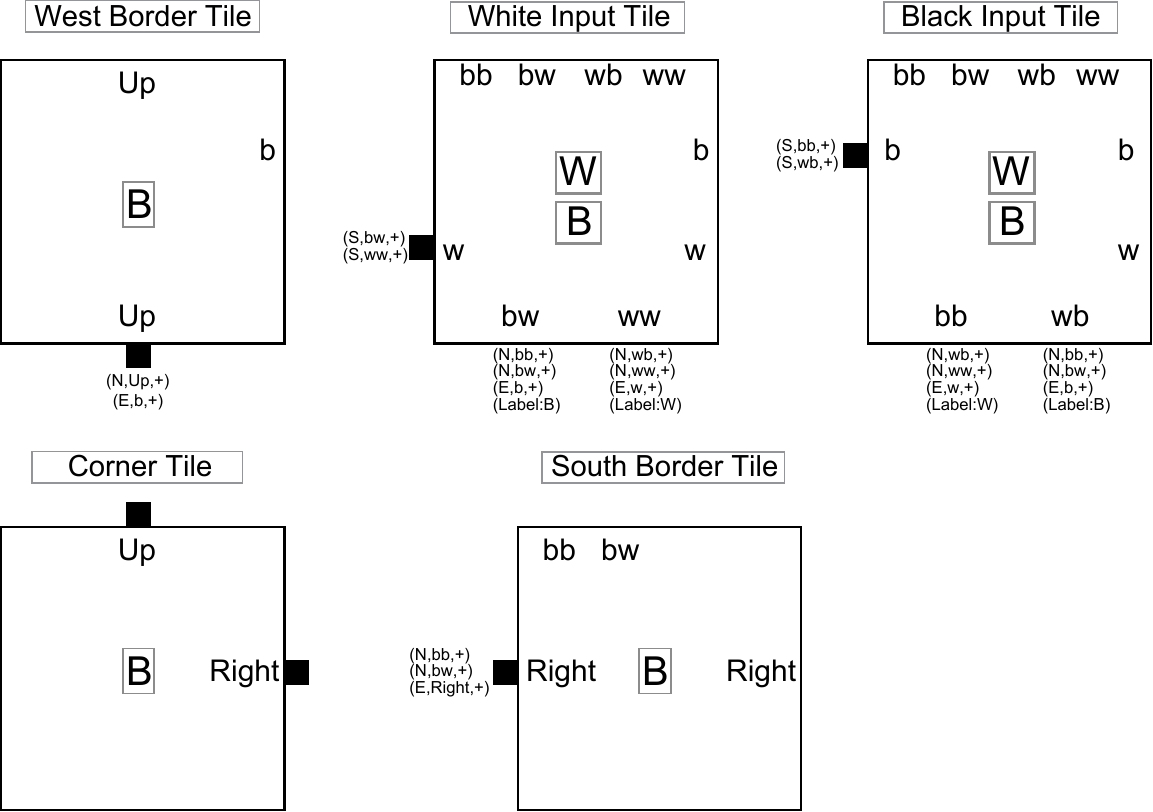}
\caption{This set of tiles weakly self-assembles the Sierpinski Triangle.}
\label{robfigs:SierpTilesTemp1}
\end{center}
\end{figure}

The weak Sierpinski STAM tile set is given in Figure~\ref{robfigs:SierpTilesTemp1}. The construction is very similar to standard aTAM constructions for weakly self-assembling the Sierpinski triangle (e.g. \cite{RoPaWi04}), with the main difference being that the aTAM construction works at temperature $\tau=2$, by having each tile (which is not on an axis) attach with two input sides.  Each input side receives as input either a $0$ or a $1$, and the value for both output sides is the \texttt{xor} of the input bits. Those tiles which output a $0$ are colored white and considered outside of the Sierpinski triangle, and those which output a $1$ are colored black and considered within it.  Since our construction works at $\tau=1$, one input direction - in this case the west - is chosen to be the first to bind.  Then, signals cause glues for either possible value of the second input to be turned \texttt{on}, allowing the tile to query the tile to its south for its value of 
0 or 1. 
Whichever of the glues answers that query then binds and activates the correct output glues as well as turning
 \texttt{on} the correct label value which identifies the tile as being either white or black.  This general signaling 
 mechanism of recruiting a new tile to the complex, activating query glues on edges that are adjacent to tiles already
 bound to the complex, and then responding to the results of the query is used throughout the temperature 1 
STAM  constructions given in this paper, and suggests the basis for simulating rectilinear systems.

\subsection{Strict Self-Assembly of the Sierpinski Triangle}
\label{sec:strict-triangle}

Our proof of Theorem~\ref{thm:strict-Sierpinski} is by construction.
Here, we provide a brief overview.

Define $f(x,y)$ as the function which takes as input a point $(x,y)$ and which returns the set of $4$ points which correspond to $(x,y)$ at a scale factor of $2$, that is, the $2 \times 2$ square of points $\{(2x+a,2y+b) \mid a,b \in \{0,1\}\}$.  (For instance, $f(1,1) = \{(2,2),(2,3),(3,2),(3,3)\}$). For notation, we will refer to the $4$ points in the set $f(x,y)$ as $f(x,y)_{00}$, $f(x,y)_{01}$, $f(x,y)_{10}$, and $f(x,y)_{11}$ with subscripts corresponding to the values for $a$ and $b$, given as $00$, $01$, $10$, and $11$, respectively.  See Figure~\ref{fig:ST-block} for a clarification of this notation.

Let $S_{2\triangle} = \{f(x,y) \mid (x,y) \in S_{\triangle}\}$ be the Sierpinski triangle at scale factor $2$, i.e. where each point in the original Sierpinski triangle is replaced by a $2 \times 2$ square of points, which we will refer to as a \emph{block}.  To prove Theorem~\ref{thm:strict-Sierpinski}, we now present an STAM system, $\mathcal{T}_{2\triangle} = (T_{2\triangle},1)$ which strictly self-assembles $S_{2\triangle}$.  At a high-level, it does so by weakly self-assembling $S_{2\triangle}$ by treating each block $f(x,y)$ as a single tile which receives one input each from the block to its south and the block to its west.  Each input is either a $0$ or $1$, and the block performs the equivalent of an $\verb"xor"$ operation on those inputs and outputs the result to its north and east.  A block $f(x,y)$ which outputs a $1$ corresponds to a point $(x,y) \in S_{\triangle}$ and thus a location which must remain tiled in the final assembly (shown as grey locations in Figure~\ref{fig:Sierpinski-triangle}).  A block $f(x,y)$ which outputs a $0$ instead corresponds to a point $(x,y) \not\in S_{\triangle}$ and must eventually be removed from the final assembly (shown as white locations in Figure~\ref{fig:Sierpinski-triangle}).  Whenever a white region is completely tiled and completely surrounded by blocks corresponding to grey positions (note that all white regions in $S_{\triangle}$ are surrounded by grey positions), glue deactivation is used to ``eject'' the blocks of that white region as a set of ``junk'' supertiles.  Those junk supertiles are then broken down into constant sized terminal supertiles (of sizes 3, 4, and 6) which are unable to attach to any portion of the infinitely growing assembly, and thus remain inert junk assemblies.

\begin{proof}[Proof sketch]

\begin{figure}[htb]
\begin{center}
\includegraphics[width=4.0in]{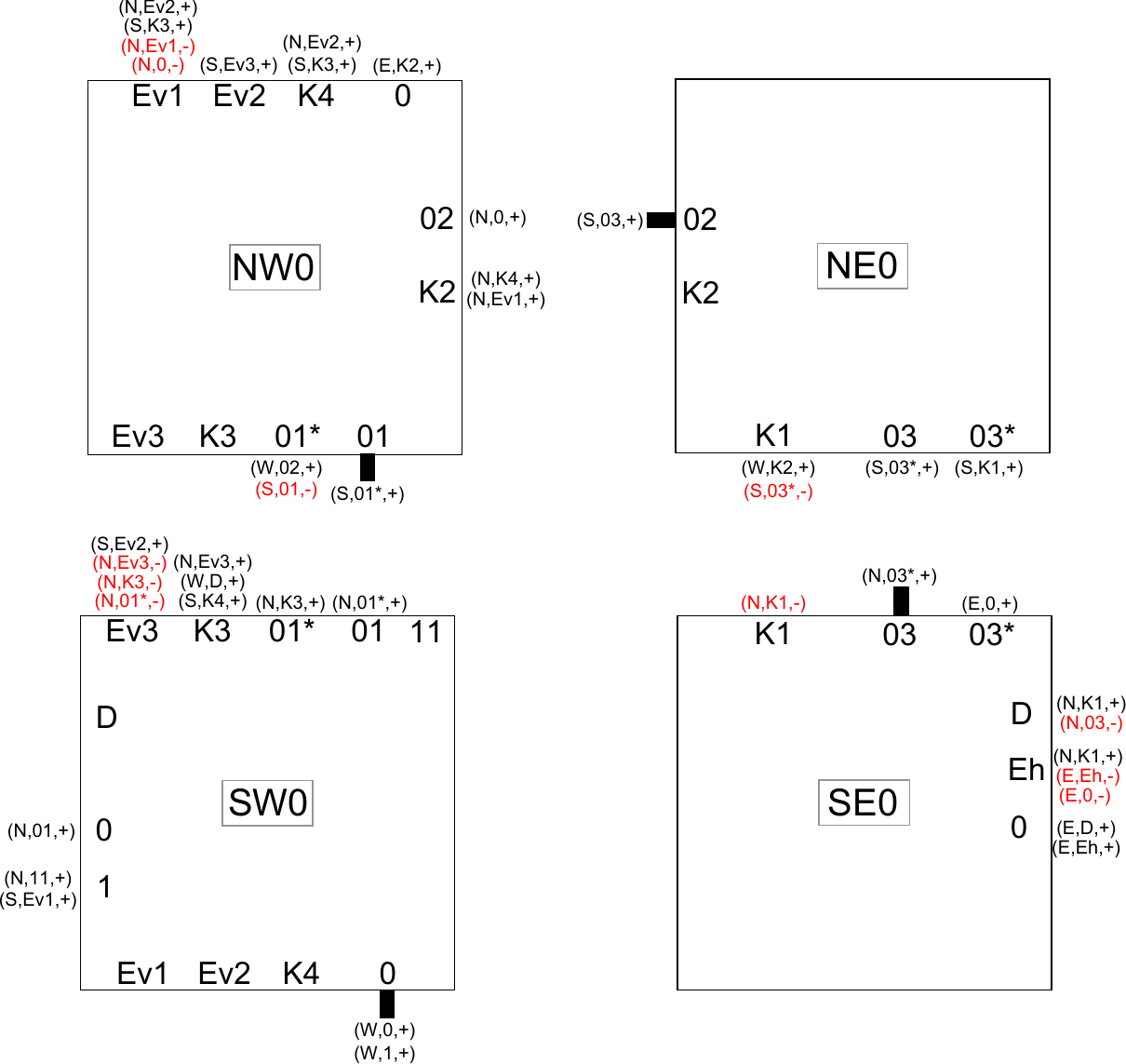}
\caption{
First group of tile types which strictly self-assembly the Sierpinski triangle}
\label{fig:ST0}
\end{center}
\end{figure}

\begin{figure}[htb]
\begin{center}
\includegraphics[width=4.0in]{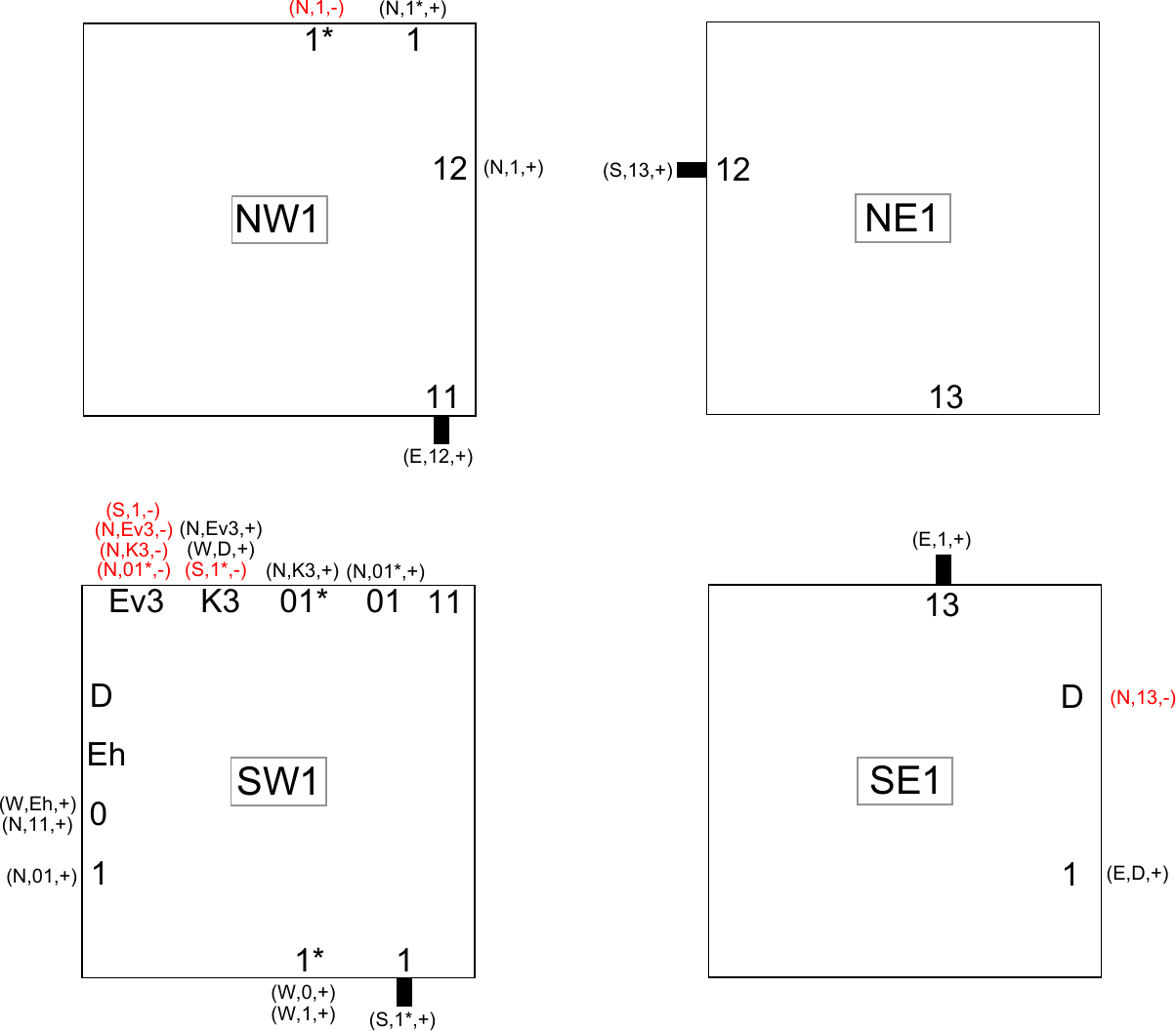}
\caption{
Second group of tile types which strictly self-assembly the Sierpinski triangle}
\label{fig:ST1}
\end{center}
\end{figure}

\begin{figure}[htb]
\begin{center}
\includegraphics[width=4.5in]{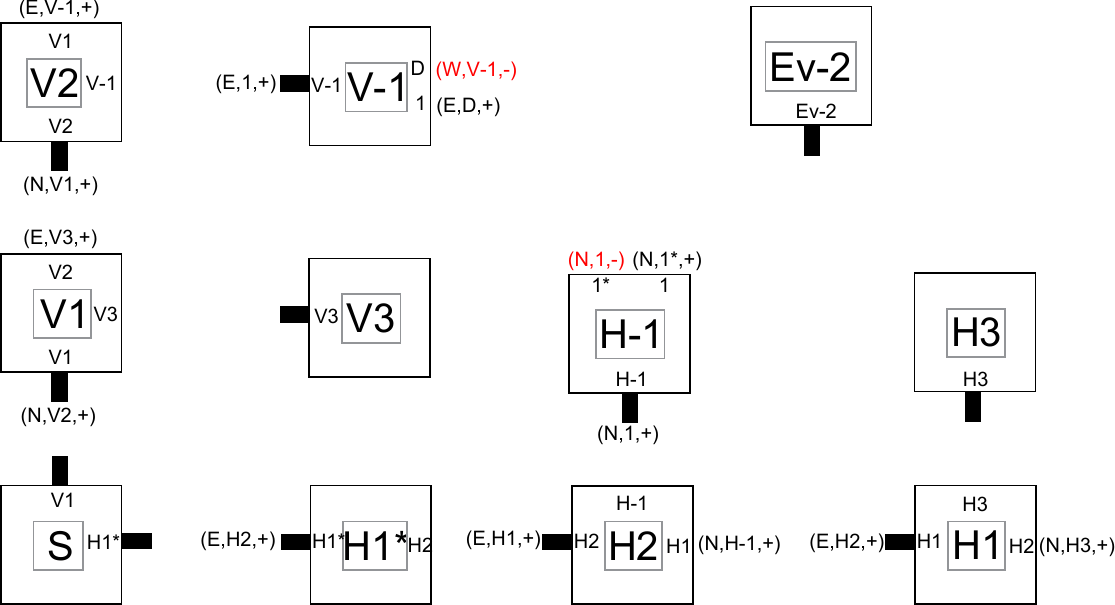}
\caption{
Third group of tile types which strictly self-assembly the Sierpinski triangle.  With the exception of the tile type labeled `Ev-2', these tile types tile the locations along the positive $x$ and $y$ axes, which are all included in $S_{\triangle}$.}
\label{fig:STx}
\end{center}
\end{figure}

The $19$ active tile types which compose $T_{2\triangle}$ are depicted in Figures~\ref{fig:ST0}, \ref{fig:ST1}, and \ref{fig:STx}. The tiles of Figures~\ref{fig:ST0} and \ref{fig:ST1} form 2 $\times$ 2 blocks
(Figure ~\ref{fig:Blocks}) that function like individual tiles in the
aTAM construction and the above construction.  Each block responds to the inputs from the south
and west, yielding an output that is the \texttt{xor} of the
binary inputs. The tiles of Figure~\ref{fig:STx} form the 2 $\times$ 2 seed block, and the horizontal and vertical blocks
along the axes, with the exception of tile Ev-2 which only binds to junk assemblies after they are ejected from the
construction, initiating the signals that further break down these junk assemblies into smaller pieces.
The seed, horizontal and vertical blocks are shown as part of the assemblies in
Figures~\ref{fig:ST-begin}-\ref{fig:ST-empty}.
Initially, based on the glues that begin in the \texttt{on} state,
only tiles of type $S$ can bind to tiles of type $V1$ and $H1$, to
begin building the seed block. These initial bindings initiate a series of glue activations which can, using tiles of the types shown in Figure~\ref{fig:STx}, tile each of the blocks corresponding to the points of the positive $x$ and $y$ axes in $S_{\triangle}$, namely the points $\{f(x,0) \mid x \in \mathbb{N}\} \cup \{f(0,y) \mid y \in \mathbb{N}\}$.  Arbitrarily large portions of the axes can form without the need for any tiles filling in the ``interior'' 0- and 1-blocks (in positions $f(x,y)$ where $x > 0$ and $y > 0$).  However, no interior block $f(x,y)$ can be fully tiled until blocks $f(x,0)$ and $f(0,y)$ have received tiles.  In fact, the first tile placed in $f(x,y)$ is always $f(x,y)_{00}$, and for interior blocks this tile always attaches first to $f(x,y-1)_{01}$ (that is, the north-west tile of the block immediately to its south). For the rest of the discussion, we will only be referring to interior blocks unless explicitly stated.

\begin{figure}[htb]
\begin{center}
\includegraphics[width=5in]{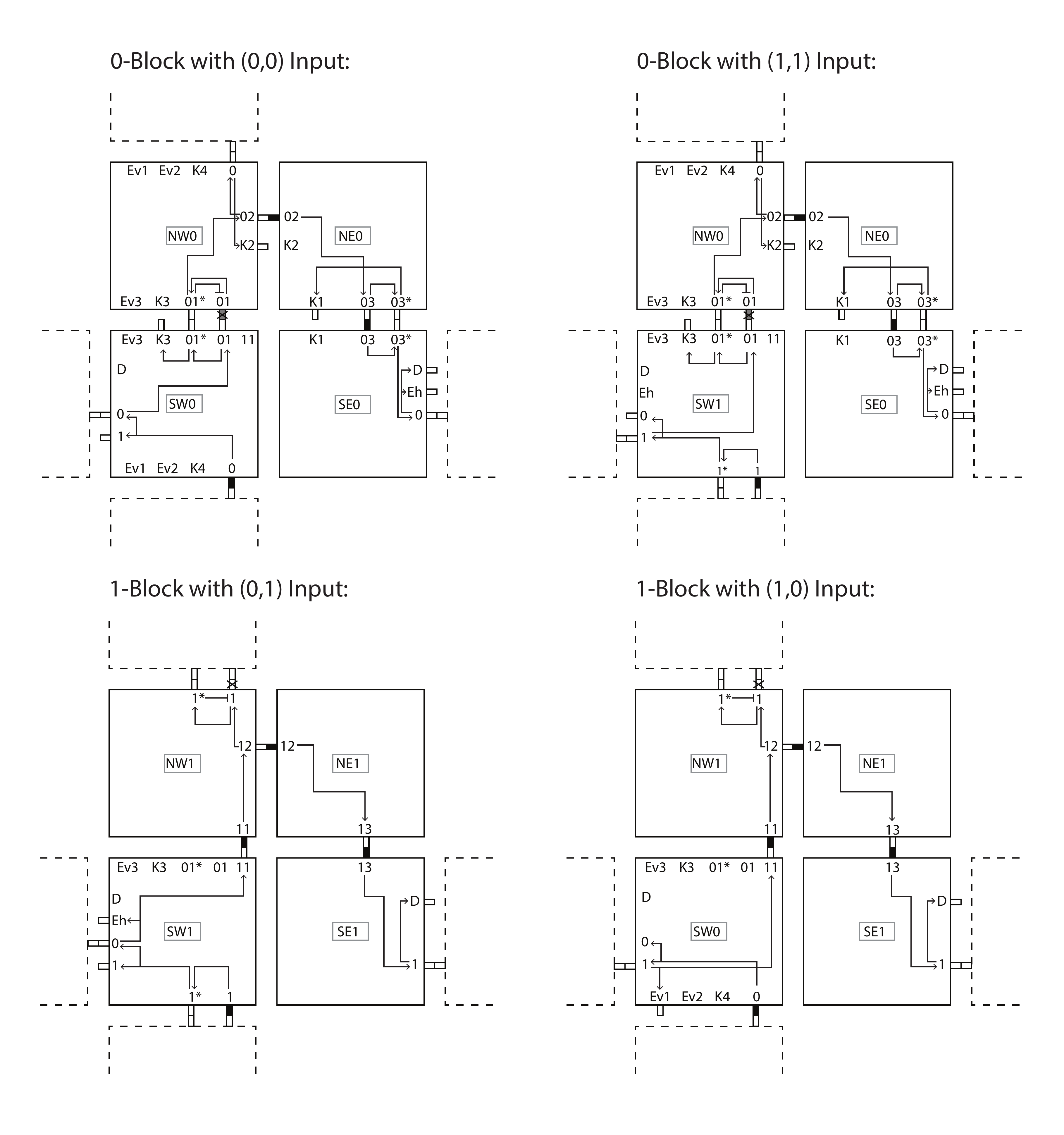}
\caption{
The interior 0- and 1-blocks formed during the strict Sierpinski construction. Signals that have been activated during
assembly of the blocks are shown. Glues that are initially active in the tile set are
colored black or grey. Glues that have been activated during the process of constructing the block are shown in
outline, those that have been deactivated are crossed out, and dashed lines are used to indicate the positions of
tiles outside the block that bind to the tiles inside the block. }
\label{fig:Blocks}
\end{center}
\end{figure}

\begin{figure}[htp]
\centering
  \subfloat[][The beginning of the formation of the axes of the Sierpinski triangle (which can continue arbitrarily far without the non-axes locations being tiled).  Since the scale factor is $2$, the $2 \times 2$ squares which are included in the pattern are shaded with a darker grey.]{%
        \label{fig:ST-axes}
        \centering
        \includegraphics[width=2.5in]{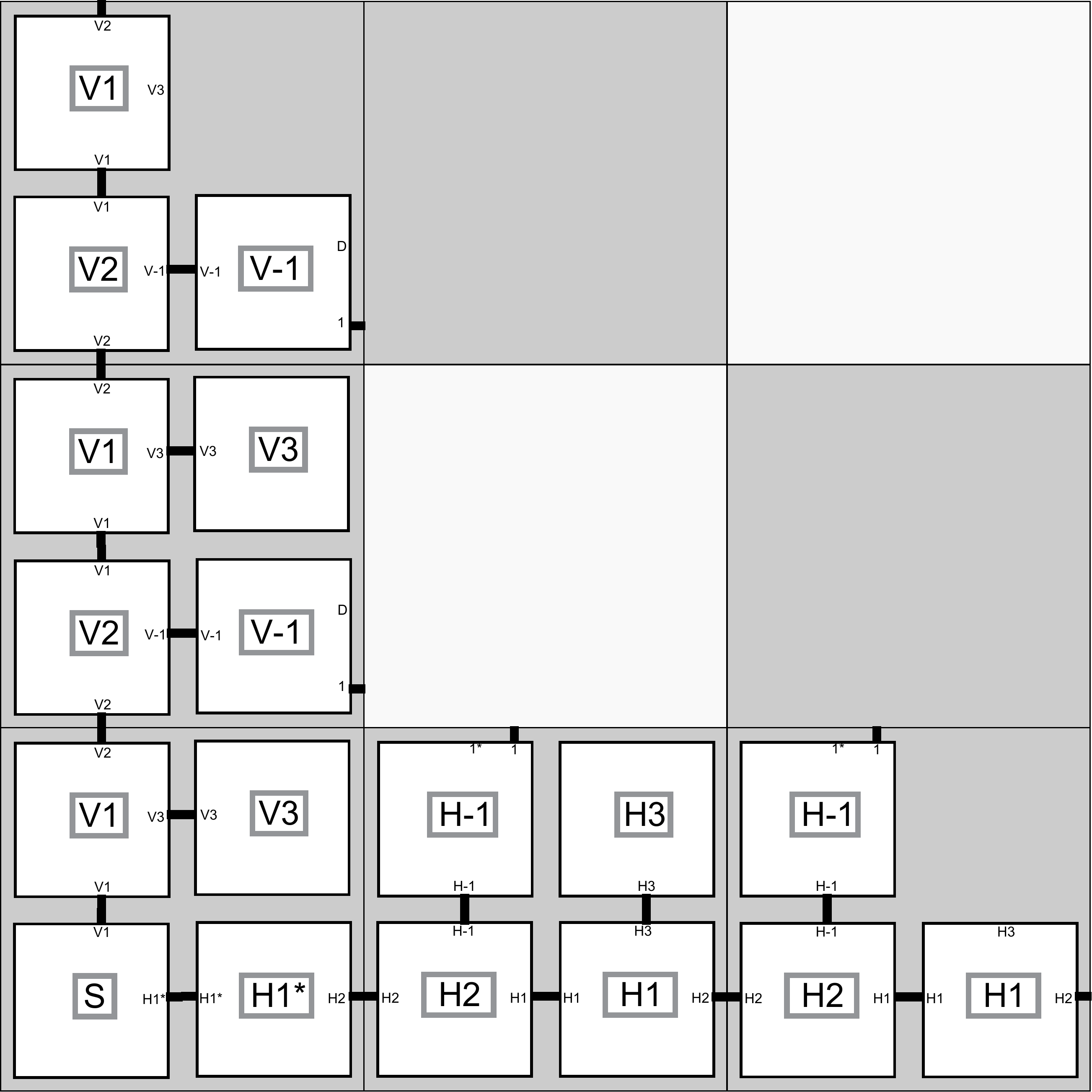}
        }%
        \hspace{25pt}
  \subfloat[][The Sierpinski triangle at the point where the first dissociation becomes possible.  The tiles in the area with the yellow background will dissociate as one ``junk'' supertile.]{%
        \label{fig:ST-first-dissoc}
        \centering
        \includegraphics[width=2.5in]{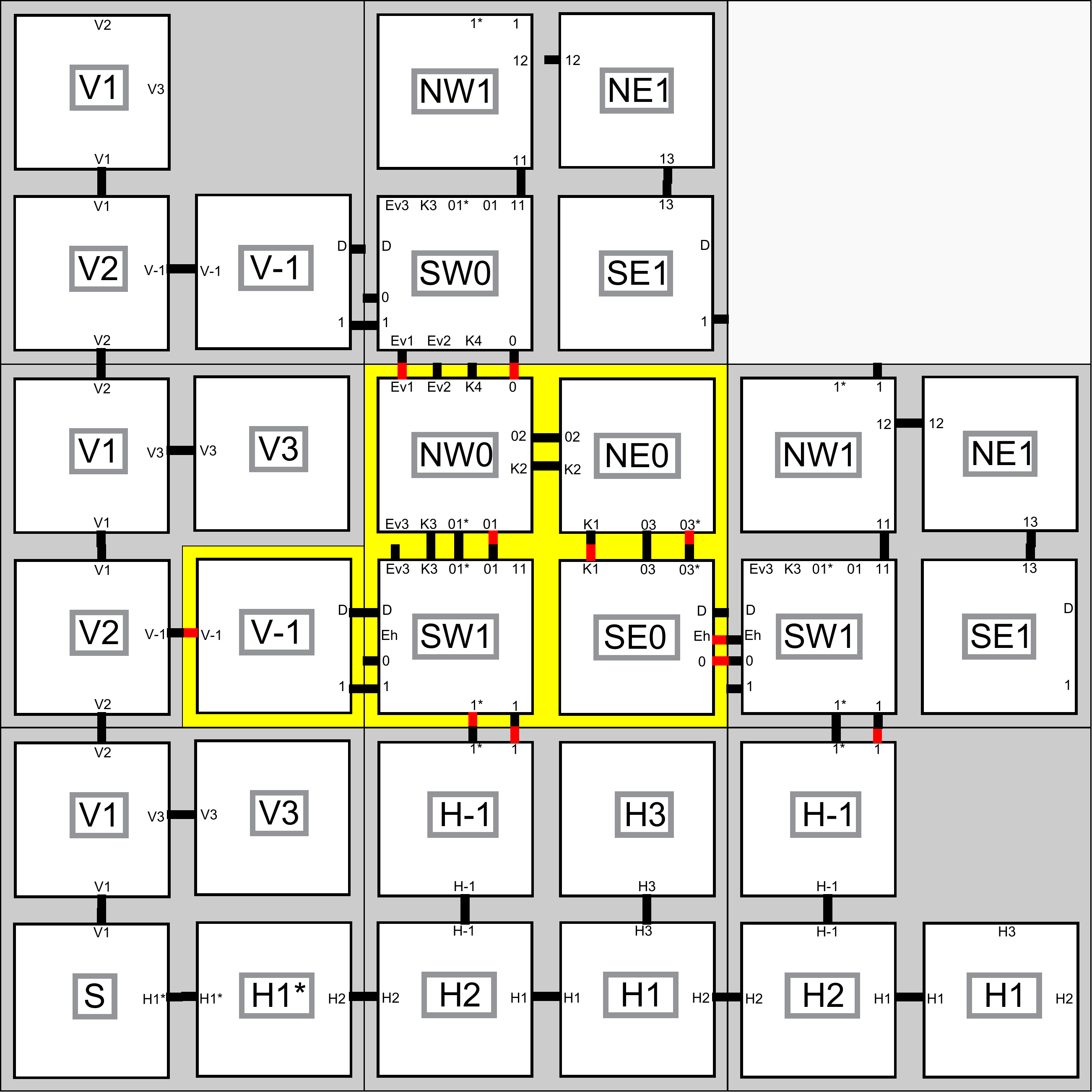}
        }%
  \centering
  \caption{The initial stages of the strict self-assembly of the Sierpinski triangle.}\label{fig:ST-begin}
\end{figure}

The first tile to attach in a 0- or 1-block is either $SW0$ or $SW1$ depending on whether or not the output from the block to the south is a $0$ or $1$, respectively (Figure ~\ref{fig:Blocks}).
(An example of the assembly process can be seen in Figures~\ref{fig:ST-begin}-\ref{fig:ST-empty}.) Once it also binds to the block to its west to receive its second input, it is able to determine the correct output for that block and activate binding glues on its north which specify the identity of that entire block as either $0$ (in a white position) or $1$ (in a grey position).  If it is a $0$-block ($1$-block), the rest of the block fills in with the $NW0$, $NE0$, and $SE0$
($NW1$, $NE1$, and $SE1$) tiles, in order.  Note that either type of block can have either a $SW0$ or $SW1$ tile in its southwest position (the $0$ and $1$ merely corresponding to the identity of the first input), since the first input is not sufficient to determine the identity of the block. The 0- and 1- blocks produce the weak Sierpinski pattern at scale factor
2 until glue deactivations begin to occur.

\begin{figure}[htb]
\begin{center}
\includegraphics[width=4.5in]{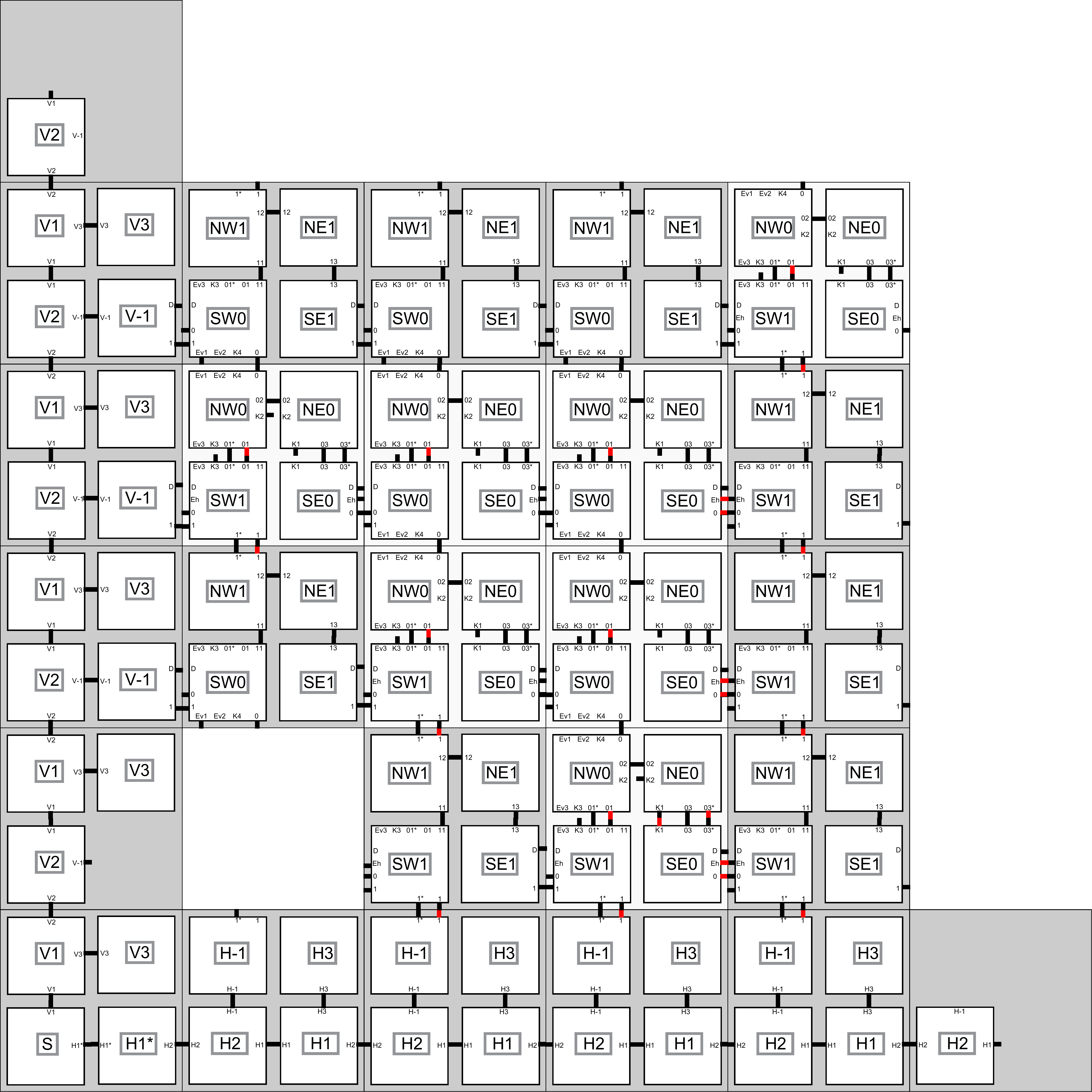}
\caption{
After weak self-assembly of the first three stages of the Sierpinski triangle plus an extra row and column to provide the necessary signals which initiate the dissociation of the interior white portions.  Those signals have only partially propagated into the white portions in this figure.}
\label{fig:ST-full}
\end{center}
\end{figure}

\begin{figure}[htb]
\begin{center}
\includegraphics[width=4.5in]{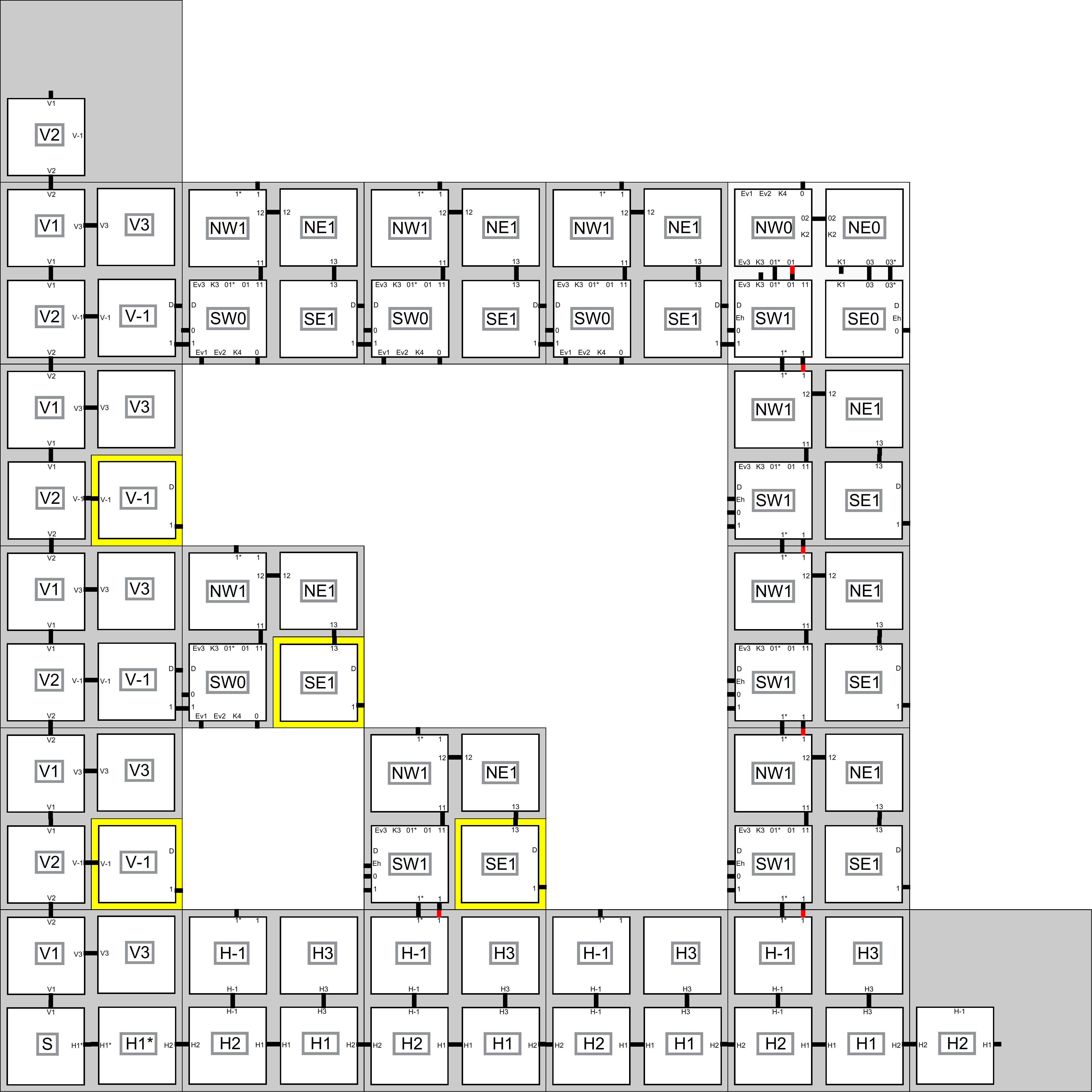}
\caption{
After weak self-assembly of the first three stages of the Sierpinski triangle and necessary dissociation.  The yellow squares indicate locations where tiles previously dissociated and new tiles have filled in.}
\label{fig:ST-empty}
\end{center}
\end{figure}

Next, we describe the dissociation process that removes the $0$-blocks from a region of the produced assembly
once that region is fully enclosed by $1$-blocks.
After blocks determine their identity as $0$- or $1$-blocks, they presenting that output to their north and east, allowing further block assembly. $1$-blocks display glues to their south or east, while $0$-blocks also activate glues on their output (north and east) sides which are used to detect that they are bordered by a $1$-block and begin the dissociation process.  When a $1$-block forms in the block $f(x,y)$, it must be the case that one of the inputs to the tile in position $f(x,y)_{00}$ is a $0$.  If $f(x,y)_{00}$ is of type $SW0$, that input was to the south, and if it is $SW1$ then that input came from the west.  Once both inputs have been determined, such a $SW0$ ($SW1$) tile will activate a glue which initiates a $Ev1$ ($Eh$) signal to the south (west).  This signal will be received by the bordering $0$-block and cause it to dissociate from the $1$-block and pass the signal further into the block and the entire white region.  The combination of $Ev$ and $Eh$ signals which pass through $0$-blocks in a white region bounded by $1$-blocks (note that they must be completely surrounded by $1$-blocks since the order of block growth is strictly up and to the right and those signals are only initiated from the south and east sides of $1$-blocks) cause them to separate into vertical chunks which can each dissociate as single supertiles.  The design of the signal propagation and glue deactivation is such that the junk supertiles can only dissociate as supertiles which are bounded on both the north and south by $1$-blocks (although they may only be $1$ block wide and thus bounded on the east and west by $0$-blocks).  This fact is extremely important to the process which further breaks apart those junk supertiles into constant sized pieces but also ensures that they will always be broken apart and never interfere with further growth of the assembly representing the Sierpinski triangle.

\begin{figure}[htb]
\begin{center}
\includegraphics[width=4.5in]{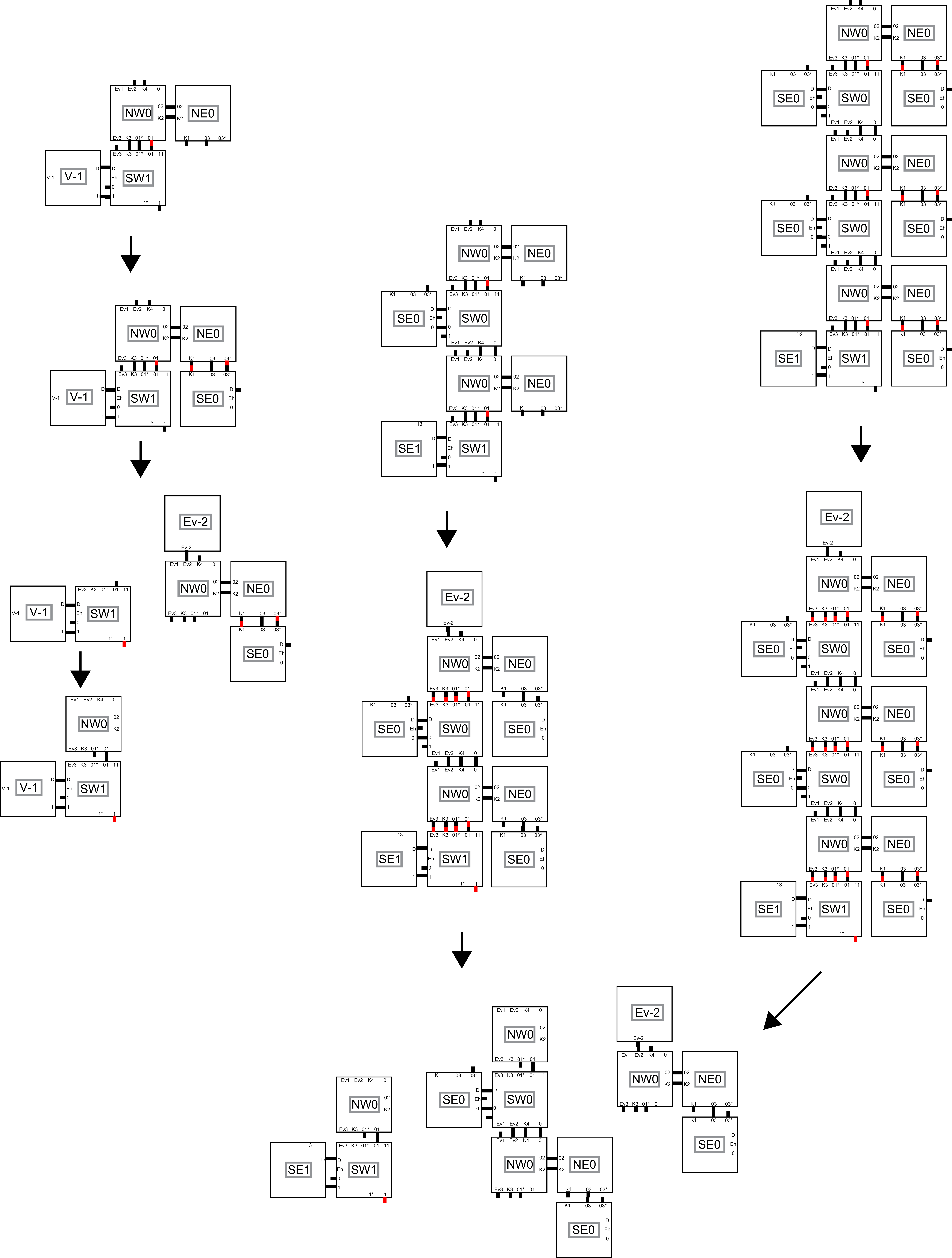}
\caption{
The ``junk'' ejected from the first three stages and the pathways along which it is broken down following the attachment of $Ev2$ tiles.  Note that at some edges, singleton tiles can attach to the junk assemblies.  However, all junk assemblies are guaranteed to break down into sizes of $3$, $4$, or $6$ and become terminal.}
\label{fig:ST-junk-digestion}
\end{center}
\end{figure}

The first phase of dissociation breaks the white regions into vertical columns which were previously bounded above and below by grey portions.  (The process by which some example junk assemblies are broken down - including some partial re-growth with the addition of singleton $NW0$ and $SE0$ tiles - into latent, constant sized junk, is shown in Figure~\ref{fig:ST-junk-digestion}.)  This means that the bonds between the white and grey regions are broken on the north and south, and also the bonds on the left and right of each column of blocks (both with white and grey blocks) are broken.  The careful design of the signals which cause the dissociation of the white regions, always beginning from grey regions to the north and east and initially only causing dissociation with bordering grey blocks to the north and south (and not other white blocks to the north and south of blocks in potentially large white regions), ensures that when junk assemblies are able to dissociate that the glues on their borders which remain active do not allow them to re-attach to any other portion of the growing assembly except for those which have no blocks formed to the north of them.  Additionally, if they are larger than $6$ tiles they are guaranteed to have an $NW0$ tile in the northwest corner of each block in their northernmost row, and that $NW0$ tile will have an active $Ev2$ glue on its north.  Since any position in which these junk assemblies can re-attach cannot have blocks to their north, the $Ev2$ glue will allow an $Ev-2$ tile to attach, and this attachment will initiate the second phase of dissociation in which a chain of signals cause the junk assembly to break up into fixed height portions.  Any re-attachment of the junk assemblies to the growing structure will therefore be temporary and also not able to cause incorrect growth by initiating invalid signals.  Thus the junk assemblies are broken into fixed width and height portions in two phases.

It is notable that junk assemblies can allow singleton $NW0$ and $SE0$ tiles to attach at some point during their break up into constant sized pieces, but that only these single attachments are possible and no additional signal activation is possible which would allow the junk assemblies to make further attachments.  Another interesting aspect of the dissociation process is that white blocks which have grey blocks bordering them on their west side will cause a single tile of those grey blocks to dissociate with them.  The newly formed hole in the grey block will be re-filled by either an $SE1$ or a $V$-$1$ tile.  The reason for this is to ``hide'' the active $0$ glue between the original $SE1$ tile and the $SW1$ tile to its east, which could allow incorrect binding of the junk assembly.  In fact, the need to hide active bonds which cannot be guaranteed to be deactivated in this asynchronous model is the reason for the complexity of the dissociation process and the size of the final, terminal junk assemblies.

Through this process in which weak self-assembly of the Sierpinski triangle proceeds until white regions are surrounded by grey blocks, and then those white regions are forced to dissociate as arbitrarily large junk assemblies which are then further broken down into constant sized junk assemblies, all the while being guaranteed not to cause incorrect assembly by binding to the still growing structure, provides the correct strict self-assembly of the Sierpinski triangle at scale 2.

\end{proof}

\subsubsection*{Acknowledgments}
The authors would like to thank Nata\v{s}a Jonoska and Daria Karpenko for fruitful discussions and comments on this work.

\bibliographystyle{amsplain}
\bibliography{tam,model_justification}

\end{document}